\documentclass[aps,pre,twocolumn,superscriptaddress,showpacs,longbibliography]{revtex4-1}
\usepackage{epsfig}
\usepackage{amsmath,amssymb,bm}
\usepackage{graphics}
\usepackage{epsfig}
\usepackage{graphicx}
\usepackage{verbatim}
\begin{document}
\title{
%%$e$-
%%Giant viable components of directed interdependent networks
Complex network view of evolving manifolds
}
\author{Diamantino~C.~da~Silva}
\affiliation{Departamento de F{\'\i}sica da Universidade de Aveiro $\&$ I3N, Campus Universit\'ario de Santiago, Aveiro 3810-193, Portugal}
\author{Ginestra~Bianconi}
\affiliation{School of Mathematical Sciences, Queen Mary University of London, London, E1 4NS, United Kingdom}
\author{Rui~A.~da~Costa}
\affiliation{Departamento de F{\'\i}sica da Universidade de Aveiro $\&$ I3N, Campus Universit\'ario de Santiago, Aveiro 3810-193, Portugal}
\author{Sergey~N. Dorogovtsev}
\affiliation{Departamento de F{\'\i}sica da Universidade de Aveiro $\&$ I3N, Campus Universit\'ario de Santiago, Aveiro 3810-193, Portugal}
\affiliation{A.F. Ioffe Physico-Technical Institute, St. Petersburg 194021, Russia}
\author{Jos\'e~F.~F. Mendes}
\affiliation{Departamento de F{\'\i}sica da Universidade de Aveiro $\&$ I3N, Campus Universit\'ario de Santiago, Aveiro 3810-193, Portugal}
%%
%%\author{Diamantino~C.~da~Silva}
%%
%%\affiliation{Departamento de F{\'\i}sica da Universidade de Aveiro $\&$ I3N, Campus Universit\'ario de Santiago, 3810-193 Aveiro, Portugal}
%%
%%\date{}
%%
%%\maketitle
\begin{abstract}

We study complex networks formed by triangulations and higher-dimensional simplicial complexes 
%%of 
representing closed evolving manifolds. In particular, for triangulations, the set of possible transformations 
%%(so called triangle mesh operations) 
of these networks is restricted by the condition that at each step, all the faces must be triangles. 
%%We show that each of these transformations 
%%%%operations 
%%can be performed in a sequence of steps, in which a single elementary transformation is applied in a specific  
%%%%different 
%%order. 
Stochastic application of these operations leads to random networks with different architectures. 
We perform extensive numerical simulations and explore the geometries of growing and equilibrium complex networks generated by these transformations and their local structural properties. This characterization includes the Hausdorff and spectral dimensions of the resulting networks, their degree distributions, and various structural correlations. Our results reveal a rich zoo of architectures and geometries of these networks, some of which appear to be small worlds while others are finite-dimensional with Hausdorff dimension equal or higher than the original dimensionality of their simplices. 
The range of spectral dimensions of the evolving triangulations turns out to be from about $1.4$ to infinity. 
Our models include simplicial complexes representing manifolds with evolving topologies, for example, an $h$-holed torus with a progressively growing number of holes. This evolving graph demonstrates features of a small-world network and has a particularly heavy-tailed degree distribution. 
%%emergent dimensionalities
\end{abstract}
%%
%%\pacs{64.60.-i ???, 05.40.-a ???, 64.60.ah ???, 64.60.F- ???}
%%
\maketitle
%%
%%

%%%%%%%%%%%%%%%
%%%%%%%%%%
%%%%%%%%%%%%%%
%%%%%%%%%%%

\section{Introduction}
\label{sec:1}

In mathematics, engineering, and various fields of natural sciences including, in particular, quantum gravity 
%%In quantum gravity 
\cite{ambjorn1997quantum,ambjorn1997geometry,rovelli2014covariant,baez2000introduction,ambjorn2010quantum,lombard2016network,kruger2016simulating}, 
%%... , ... , and many other fields of natural sciences and mathematics, 
%%physics, 
manifolds 
%%(topological spaces, locally homeomorphic, i.e. similar, to Euclidean spaces) 
play a pivotal role. 
The manifolds are topological spaces locally homeomorphic 
%%, i.e., similar, 
to Euclidean spaces \cite{edelsbrunner2010computational}. 
Simple examples of manifolds are the circle, the sphere, the torus, etc. The 
%%natural 
discrete construction of manifolds is by simplices (triangle: two-simplex; tetrahedron: three-simplex, etc.), where a simplex is a building block, 
%%of a manifold, 
and a simplicial complex is homeomorphic to a manifold. The simplicial complex construction is possible for, in particular, any smooth (in other words, differentiable) closed manifold. For example, complexes of triangles (triangulations) are homeomorphic to various surfaces, including the sphere, the torus, etc. 
Simplicial complexes 
%%can be 
are extensively treated and used as the discrete version of manifolds \cite{munkres1984elements,kong1989digital,higuchi2001combinatorial,keller2011curvature}. 
One should consciously note the following points: (i) Other discrete versions of manifolds, not based on simplices, are also possible, e.g., various grids. (ii) We consider simplicial complexes constructed of only  simplices of equal dimension. (iii) The simplicial complex based discrete description of an arbitrary manifold should include the full set of the edges lengths of the simplices. As is natural, simplicial complexes with all edges equal can represent only a small fraction of manifolds.    
It was recently proposed to treat simplicial complexes as complex networks 
formed by the vertices and edges of the simplices \cite{wu2015emergent,bianconi2015complex,bianconi2015b-complex,bianconi2016emergent,bianconi2016network,courtney2016generalized}, and so to tackle the problem of evolving manifolds by using apparatus and models taken from the theory of evolving complex networks. This treatment conforms well to the modern interest in properties of networks embedded into various metric spaces \cite{papadopoulos2012popularity,zuev2015exponential,krioukov2016clustering}. 

The works \cite{bianconi2015complex,bianconi2015b-complex,bianconi2016emergent,bianconi2016network,wu2015emergent,courtney2016generalized} considered simplicial complexes representing growing manifolds with a border. Notably, the growth in the evolution models proposed in these papers was wholly due to the attachment of new elements (simplexes or their parts) to the border of a simplicial complex. After the attachment, the resulting part of the simplicial complex did not evolve. 
%%evolved in the proposed growth models. .... due to progressive addition of new elements (simplicial complexes or their parts). 
%%For the rationale behind the attention to manifolds with a border see Regge calculus 
%%Discrete manifolds of dimension $d$   that are topologically isomorphic to a $d$-dimensional sphere, i.e. with a border can be simply reduced to $d-1$ manifolds without border topologically isomorphic to a $d-1$ spherical surface  as can be seen for instance in the framework of Regge theory 
Note however that discrete manifolds of dimension $d$ that are topologically
isomorphic to a $d$-dimensional sphere having all their nodes laying on the border, like the ones studied in Refs. \cite{bianconi2015complex,bianconi2015b-complex,bianconi2016emergent,bianconi2016network,wu2015emergent,courtney2016generalized}, can be simply reduced to $(d-1)$-dimensional manifolds without border
topologically isomorphic to a $(d-1)$-dimensional spherical surface as
can be seen for instance in the framework of Regge theory \cite{dittrich2012canonical}.   
In the present article we explore a large variety of 
simplicial complexes ``triangulating'' 
evolving manifolds, including manifolds with 
%%non-trivial, 
varying topology. 
%%a principally different type of evolving manifolds. 
In particular, we consider two-simplicial complexes triangulating surfaces homeomorphic to a sphere and $h$-holed (genus-$h$) torus with a growing number of holes (genera, in other words). 
We propose a set of basic complex network models for simplicial complexes representing evolving manifolds without borders, so called closed manifolds. In this evolution, the entire manifold progressively evolves, and any part of it has a chance to be modified at any instant. These manifolds can be growing, equilibrium, and decaying, as  particular cases of evolving complex networks \cite{dorogovtsev2000scaling}, and we consider the first two of these cases in detail. 
 
We show that 
evolutionary models of this kind generate 
%this kind of evolutionary models generate 
manifolds with a wide spectrum of space dimensions, including 
%%distinct 
different from their original dimensionalities, with a small-world geometry (i.e., infinite dimensional) only as a particular case. 
This wide set of different generated Hausdorff dimensions, from $2.5$ to $\infty$, of our evolving triangulations is in sharp contrast to random planar graphs, which are known to be four dimensional typically \cite{ambjorn1997quantum}. It turned out that the range of the spectral dimensions of these triangulations is from about $1.4$ to $\infty$. 
Apart from the emergent 
%%dimensionality, 
dimensions, we describe local structural characteristics of these evolving simplicial complexes, the simplest of which are degree distributions and degree--degree correlations. 
%%Furthermore, 
Advantageously, our models can generate evolving topologies. That is, such networks in different instants of evolution can be not homeomorphic to each other.
%%Here we mean the following. 
For example, when 
%%When 
implementing our rules of evolution, 
%%with time 
a sphere can turn into a torus, then into a two-holed torus, and so on. We describe evolving manifolds homeomorphic to an $h$-fold ($h$-holed) torus, in which the number of such topological features 
(holes)  
%%``wormholes'' 
grows with time, which may be treated as a toy model of the evolving Universe. In our models, the 
holes 
%%wormholes 
emerge with higher probability in places with higher curvature, i.e., near hubs, and, in their turn, while emerging, the 
holes 
%%wormholes 
produce vertices of high degrees, and so increase curvature. In this respect, these topological 
features 
%%---wormholes--- 
are associated with hubs and co-evolve.  
%%This evolution process can be treated as a toy model of the Universe.   

%%For demonstration purposes, 
For demonstration purposes and simplicity, we mostly focus on two-manifolds, that is, surfaces and their triangulations (two-simplicial complexes). 
%%Apart from analytically found degree distributions, 
Our conclusions are mainly based on the results of extensive numerical simulations of a 
%%representative 
set of 
%%the 
evolution models of growing and equilibrium simplicial complexes, demonstrating various Hausdorff and spectral dimensions.  
For 
%%growing 
%%triangulations and 
growing complexes of two-  and higher-dimensional simplexes, we analytically obtain degree distributions. 
%%One should note that the models of the present paper is more difficult for analytical treatment than of Refs.~\cite{xxx,xxx,xxx}, so mostly simulations, but .... 

%%advantage of recursive graphs

The paper is organized as follows. 
For the sake of clarity, 
Sec.~\ref{s2} reminds of the basics on triangulations in application to closed surfaces. 
Section~\ref{s3} considers the local transformations of triangulations that keep them within the complete set of triangulations 
%%and in this sense do not spoil them 
(triangular mesh operations).
In Sec.~\ref{s4} we introduce a set of evolution models for simplicial complexes. 
%%the evolution rules for our models and describe relations between them. 
In Sec.~\ref{s5} we describe our findings concerning local features of the resulting simplicial complexes, foremost, their degree distributions. Section~\ref{s6} reveals key metric properties of evolving triangulations, namely emergence of higher space dimensions in these systems and their spectral dimension.  Section~\ref{s7} describes evolving simplicial complexes representing 
manifolds whose topology evolves with time. Finally, in Sec.~\ref{s8} we discuss our results and their consequences.

%%  triangulations simplicial complices   

%%%%%%%%%%%%%%%
%%%%%%%%%%
%%%%%%%%%%%%%%
%%%%%%%%%%%
  
\section{Triangulations of closed surfaces}
\label{s2} 

%%We particularly focus on triangulation networks representing closed surfaces. 
For the sake of clarity, 
here we remind of a few features of triangulation networks (complexes of triangular faces) representing closed surfaces which are of particular interest in this study. 
%%The point is to explain 
We focus on topological invariants and on a local curvature for such triangulations and the closed surfaces that these specific complexes triangulate. In Sec.~\ref{s7}, we shall consider an evolution model in which these invariants vary in time. 

The absence of a boundary in a surface has two immediate consequences for its triangulations. 
First, each edge in such a triangulation has exactly two adjacent 
%%triangles (triangular faces), 
triangular faces, and each 
%%triangle 
face has three 
%%neighboring 
edges, so we arrive at the following relation between the total number of faces $F$ and the total number of edges $E$: 
\begin{equation}
3F = 2E 
.
\label{10}
\end{equation}
Let us recall the renowned Euler formula for general polyhedra, 
\begin{equation}
\chi = F + N - E 
.
\label{20}
\end{equation}
%%
%%%%\blue{Gin: does this remain true also when the topology is non trivial? Not clear to me }
%%
Here $N$ is the total number of vertices and $\chi$ is the Euler characteristic, a topological invariant $\chi = 2(1-h)$, where $h$ is the number of holes 
piercing this polyhedra 
%%(so one can call them ``wormholes'' as in cosmology, or genera, as in topology).  
(genera, in the language of topology). 
%%(we will call them ``wormholes'' as in cosmology) 
%%in the surface. 
%%in this polyhedra and, equally, in the closed manifold that this polyhedra maps (represents as its discrete version). 
Equally, $h$ is the number of holes 
piercing the closed manifold that this polyhedra maps (represents as its discrete version). 
For a surface homeomorphic to a sphere, $\chi=2$, for a torus, $\chi=0$, and so on. 
For the $h$-holed torus, $\chi = 2(1-h)$, and the Betti numbers \cite{munkres1984elements} are $b_0=1$, $b_1=2h=2-\chi$, and $b_2=1$. 
Here the holes in the polyhedra corresponds to the holes in the $h$-holed torus that this polyhedra represents. 
%%Then, t
Taking into account Eq.~(\ref{10}), we readily get simple expressions valid for 
%%Euler's characteristic 
triangulations of closed surfaces: 
\begin{equation}
\chi =  N - \frac13 E = N - \frac12 F
.
\label{30}
\end{equation}

The second consequence of the absence of boundaries is that for any vertex, its vertex degree $q$ coincides with the number of triangles (triangular faces) $t$ attached to this vertex, $q_i=t_i$ for vertex $i$. Applying this constraint to the formula for the local curvature $R_i$ of a triangulation constructed of triangles of equal length 
\begin{equation}
R_i = 1 - \frac12 q_i + \frac13 t_i
,
\label{40}
\end{equation}
see Refs.~\cite{higuchi2001combinatorial,knill2012index,keller2011curvature}, gives for the local environment of vertex $i$ in a closed surface the following local curvature \cite{wu2015emergent}: 
\begin{equation}
R_i = 1 - \frac16 q_i
.
\label{50}
\end{equation}
Thus for such triangulations of closed surfaces, local curvature for any vertex is completely determined by its degree, and in this sense it is a secondary notion here. 
If the degree is below $6$ (i.e., $q_i=3$, $4$, or $5$), the local curvature is positive, if above $6$, then negative. 
In triangulations with boundaries, Eq.~(\ref{50}) is violated for vertices on a boundary. The degree of a vertex of this type exceeds the number of triangles attached to the vertex, and an edge on a boundary belongs only to one triangle. 

We emphasize that Eqs.~(\ref{40}) and (\ref{50}) for a local curvature are valid only if all triangular faces in a triangulation network have edges of the same length. The rules of the evolution models in this work do not include edge lengths, and we assume the edge length equality only when treating our results in terms of curvature.

%%[20] Keller, M., \& Norbert P., Cheeger constants, growth and spectrum of locally tessellating planar graphs. Mathematische Zeitschrift 268, 871 (2011). 

%% Gromov, M. Hyperbolic groups ( Springer, New York,1987). 

%%%%%%%%%%%%%%%
%%%%%%%%%%
%%%%%%%%%%%%%%
%%%%%%%%%%%

\section{Triangular mesh operations}
\label{s3}

%%%%%%%%%%%%%%%%%%%%%%%%%%%%%%%%%%%%%%%%%%%%
%%%%%%%%%%%%%%%%%%%%%%%%%%%%%%%%%%%%%%%%%%%%
%%
\begin{figure}[t]
\begin{center}
\includegraphics[scale=0.20]{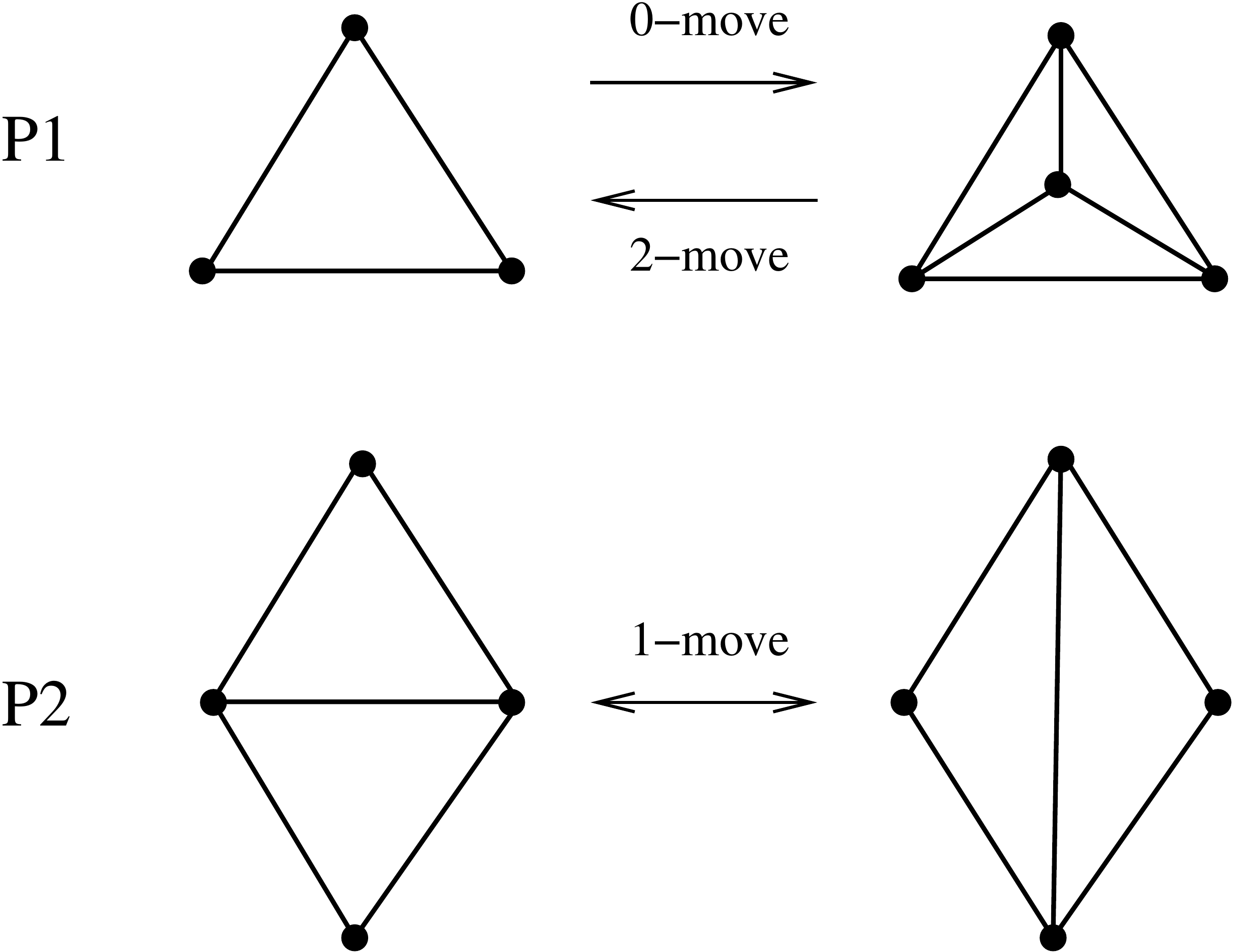}
\end{center}
\caption{
Pachner moves. 
The zero-move increases the number of faces by $2$. The two-move reduces their number by 2. 
}
\label{f1}       
\end{figure}
%%
%%%%%%%%%%%%%%%%%%%%%%%%%%%%%%%%%%%%%%%%%%%%
%%%%%%%%%%%%%%%%%%%%%%%%%%%%%%%%%%%%%%%%%%%%

%%%%%%%%%%%%%%%%%%%%%%%%%%%%%%%%%%%%%%%%%%%%
%%%%%%%%%%%%%%%%%%%%%%%%%%%%%%%%%%%%%%%%%%%%
%%
\begin{figure}[t]
\begin{center}
\includegraphics[scale=0.19]{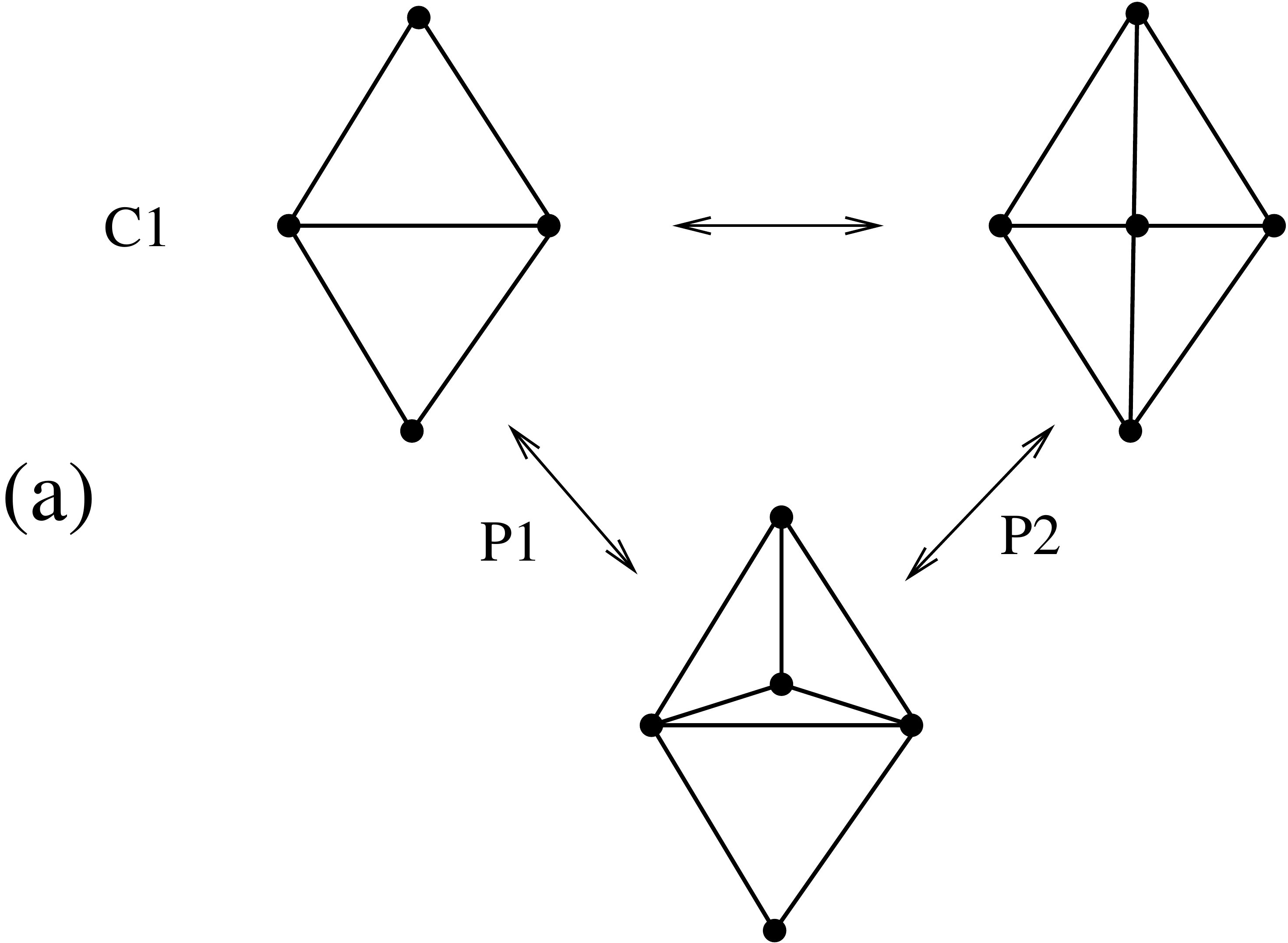}
\\[12pt]
\includegraphics[scale=0.19]{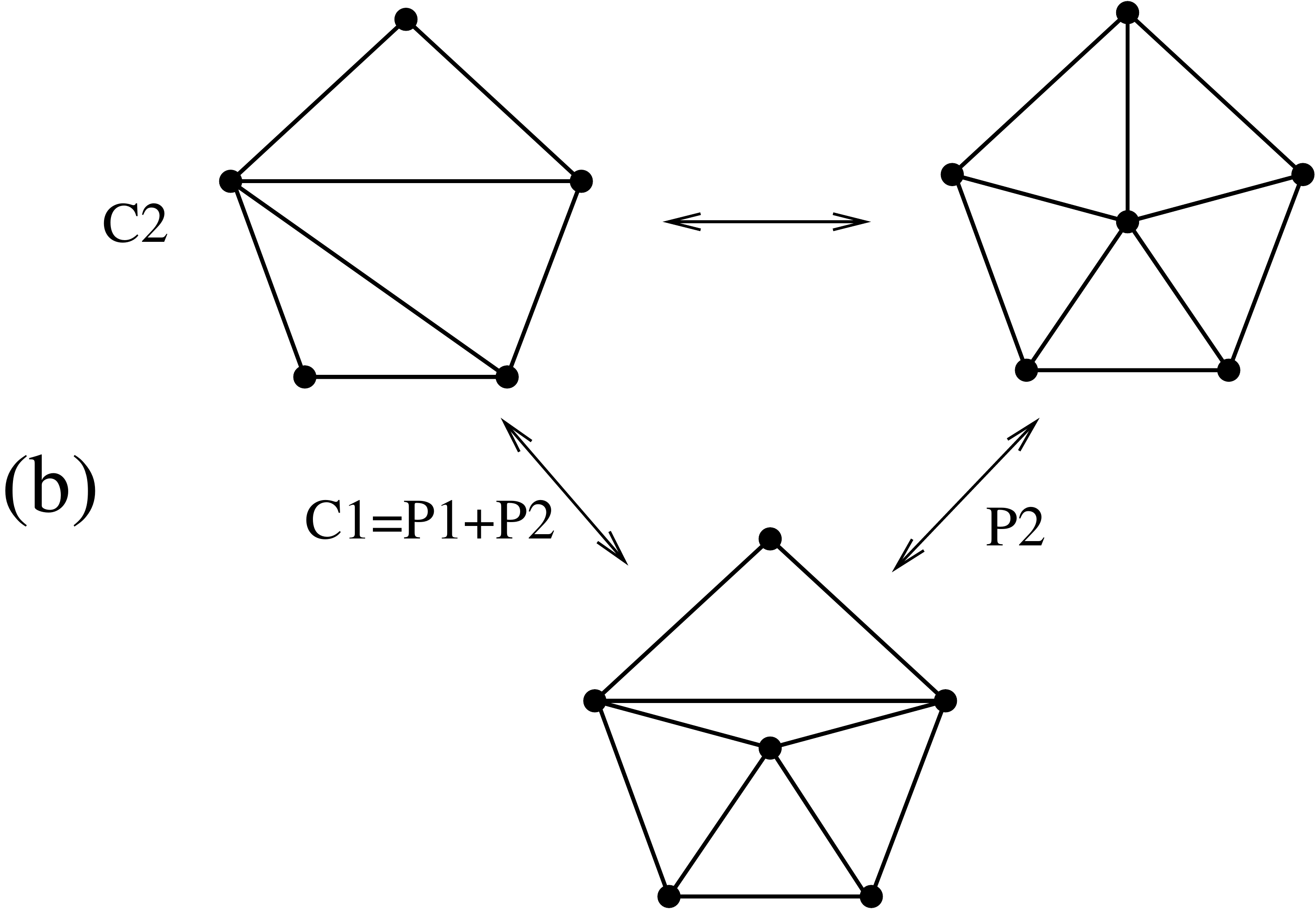}
\end{center}
\caption{
Examples of transformations that can be reduced to a sequence of Pachner moves. 
(a) 
%%Operation C1, 
In any of two directions, operation $C1$ can be performed as a sequence two Pachner moves; from left to right: first $P1$ and then $P2$; from right to left: first $P2$ and then $P1$.  
(b) In either of two directions, operation $C2$ can be performed as a sequence of three Pachner moves.
}
\label{f2}       
\end{figure}
%%
%%%%%%%%%%%%%%%%%%%%%%%%%%%%%%%%%%%%%%%%%%%%
%%%%%%%%%%%%%%%%%%%%%%%%%%%%%%%%%%%%%%%%%%%%

%%%%%%%%%%%%%%%%%%%%%%%%%%%%%%%%%%%%%%%%%%%%
%%%%%%%%%%%%%%%%%%%%%%%%%%%%%%%%%%%%%%%%%%%%
%%
\begin{figure}[t]
\begin{center}
\includegraphics[scale=1.0]{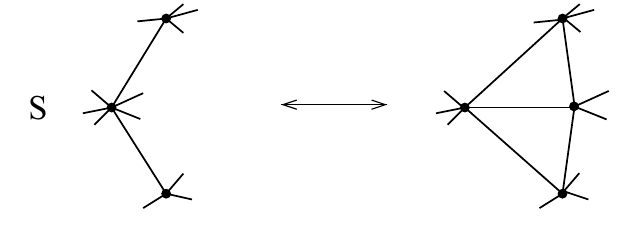}
%%{evolving_manifolds-f3}
\end{center}
\caption{
The ``elementary'' 
%%transformations 
operation $S$: splitting and merging of adjacent edges and their joint vertex. 
%%(Alexander moves \cite{aste1998dynamical,dittrich2015flux}). 
The two adjacent edges and their joint vertex are transformed into two adjacent triangles sharing 
the new edge between the joint vertex and its counterpart. 
The move from left to right creates two new faces. The move from right to left eliminates two faces. All other faces in this triangulation stay intact. 
 Any triangular mesh transformation 
%%transformation remaining a triangulation to be a triangulation (CHECK GRAMMAR! - conserving triangulability ??? ))))))) ) 
can be reduced to a finite sequence of steps, each of which is this operation.
}
\label{f3}       
\end{figure}
%%
%%%%%%%%%%%%%%%%%%%%%%%%%%%%%%%%%%%%%%%%%%%%
%%%%%%%%%%%%%%%%%%%%%%%%%%%%%%%%%%%%%%%%%%%%

%%%%%%%%%%%%%%%%%%%%%%%%%%%%%%%%%%%%%%%%%
%%%%%%%%%%%%%%%%%%%%%%%%%%%%%%%%%%%%%%%%%
%%
%%
\begin{figure}[t]
\includegraphics[scale=1.0]{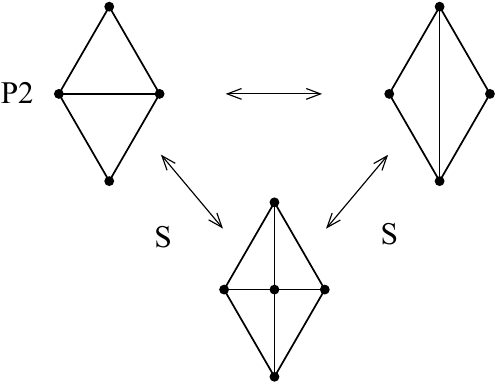}
%%{evolving_manifolds-f4}
\caption{
Pachner move $P2$ can be performed in two steps by applying $S$ twice. }
\label{f4}      
\end{figure}
%%
%%%%%%%%%%%%%%%%%%%%%%%%%%%%%%%%%%%%%%%%%
%%%%%%%%%%%%%%%%%%%%%%%%%%%%%%%%%%%%%%%%%

%%%%%%%%%%%%%%%%%%%%%%%%%%%%%%%%%%%%%%%%%%%%
%%%%%%%%%%%%%%%%%%%%%%%%%%%%%%%%%%%%%%%%%%%%
%%
\begin{figure}[t]
\begin{center}
\includegraphics[scale=1.0]{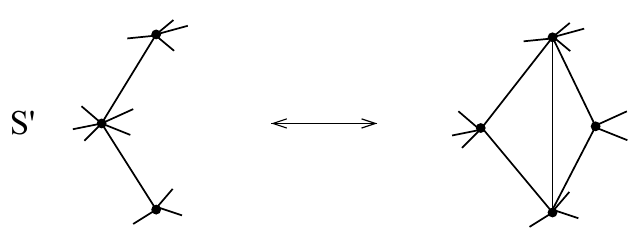}
%%{evolving_manifolds-f5}
\end{center}
\caption{
Operation $S'$. In contrast to $S$ in Fig.~\ref{f3}, the new edge here interconnects the opposite ends of the adjacent edges. Note the following restriction: the adjacent edges on the left cannot belong to the same triangle.}
\label{f5}       
\end{figure}
%%
%%%%%%%%%%%%%%%%%%%%%%%%%%%%%%%%%%%%%%%%%%%%
%%%%%%%%%%%%%%%%%%%%%%%%%%%%%%%%%%%%%%%%%%%%

Let us introduce transformations, which we use in the next section as elementary steps of the models  generating evolving triangulations, and describe relations between these transformations.  
For triangulations, the evolution rules of our models should satisfy the following conditions. At each step, all the faces must be triangles. 
It is also demanded that these transformations do not change topological features of the triangulated surface, namely the number of holes (genera) in it.   
The Pachner moves (bistellar flips) \cite{pachner1991pl,bjorner2000simplicial,cortes2002diagonal}, Fig.~\ref{f1}, are usually treated to be the minimal necessary set of these ``triangular mesh operations'':
%%The minimal necessary set of these ``mesh operations'', based on faces---triangles, is provided by the Pachner moves (bistellar flips) \cite{pachner1991pl,bjorner2000simplicial,cortes2002diagonal}, see Fig.~\ref{f1}, 
$P1$ (the so-called zero-move in one direction, creating a new vertex with three edges within a triangle, and the two-move in the opposite direction) and $P2$ (flip of a joint edge between two triangles---the one-move), in total, three moves. 
Here we focus on triangulations, but one should note that the Pachner moves are also defined for higher-dimensional simplicial complexes \cite{pachner1991pl}. 
If we consider the closed triangulated two-dimensional manifold as the boundary of a three-dimensional manifold formed tetrahedra and isomorphic to a sphere, the Pachner moves can be also interpreted as the result of gluing a single tetrahedra. Specifically the $P1$ move corresponds to gluing the new tetrahedra to a single triangular face of the boundary of the three-dimensional manifold, the $P2$ move corresponds to gluing the new tetrahedra to two incident triangular faces of the three-dimensional manifold.  
All other transformations between homeomorphic (topologically equivalent) triangulations can be equivalently performed by a sequence of Pachner moves. 
Some of such operations were shown in Refs.~\cite{loop2000managing,izmestiev2015simplicial}. 
For example, the operation $C1$ shown in Fig.~\ref{f2}(a) can be obtained in two steps by the consecutive application of two Pachner moves, $P1$ and then $P2$.   
Transformations $P1$ and $C1$ are called the Alexander moves or star subdivisions \cite{aste1998dynamical,dittrich2015flux}. 
The operation $C2$ shown in Fig.~\ref{f2}(b) can be obtained in three Pachner steps and so on.
%%One can add extra operations to this set \cite{loop2000managing,izmestiev2015simplicial}, e.g., the transformation shown in Fig.~\ref{f2}(a). This figure demonstrates, however, that this transformation can be obtained in two steps by the consiquitive application of two Pachner moves, P1 and then P2. 
%%The same is true for any other transformation of this kind, including the one shown in Fig.~\ref{f2}(b), though the equivalent sequence of Pachner moves becomes longer.   

Let us consider now the operation $S$ explained in Fig.~\ref{f3}, that is splitting of two adjacent edges and the vertex between them in such a way that the vertex and its newborn counterpart are interconnected by a new edge. This operation has two directions---two moves.  The Alexander moves $P1$ and $C1$ \cite{aste1998dynamical,dittrich2015flux} are particular cases of this operation for the situations in which the two split edges belong to the same or two neighboring triangles, respectively. Clearly, this operation can be performed in a finite number of Pachner moves, the same as in the examples shown in Fig.~\ref{f2}. 
Indeed, operation $S$ actually coincides with operations of the type shown in Fig.~\ref{f2}. 
%%Notably, 
On the other hand, in its turn, each of the Pachner moves can be performed in one or two steps of the  operation $S$. Indeed, operation $P1$ is 
%%simply 
a particular case of operation $S$, which means a single application of $S$. 
%%(Operation S1 is another particular case of S1.) 
Further, Fig.~\ref{f4} demonstrates how Pachner move $P2$ can be performed in two steps by applying $S$. 

Note that operation $S'$ introduced in Fig.~\ref{f5} (new edge interconnects the opposite ends of the adjacent edges) can be directly obtained from operation $S$ by Pachner move $P2$, and so operation $S'$ can be performed in a finite number of $S$ steps. The opposite is, however, not true, although, at first sight, operation $S'$ looks rather similar to operation $S$. The actual reason is that, as Fig.~\ref{f5} shows, operation $S'$ is not applicable to a pair of adjacent edges if they belong to the same triangle. That is, operation $S'$ is possible only in a restricted set of situations and so it in principle cannot be an ``elementary triangular mesh operation''. 

%%Finally, one can easily show that operations S1 and S2 can be performed one through the other. Indeed, they are connected by Pachner move P2, which, as we have shown can be performed by application of S1.

We conclude that any transformation of triangulation that preserves topology (number of holes in the triangulated surface) can be finally reduced to a sequence of operations $S$. So we can reduce the minimal set of triangular mesh operations to the single elementary operation $S$ (two moves) instead of three Pachner moves.  
%%Thus we described relations between operations P1, P2, S, and S$'$, which we use in the next section to generate evolving triangulations. 

%%*******
%%
%%4 and 5-cores ???? ask Rui    - what does not evolve .... ????? initial conditions ****** stacked evolution 
%%
%%Rui, Diamantino, please write this paragraph. Maybe, it will go to another place. 
%%
%%*******

%%%%%%%%%%%%%%%
%%%%%%%%%%    s4
%%%%%%%%%%%%%%
%%%%%%%%%%%

\section{Rules of evolution}
\label{s4} 

Our evolution models are organized in the following way. At each time step, 
(i) an element or neighboring elements of the simplicial complex under consideration are chosen 
%%uniformly at random or 
with some preference or, in the simplest particular case, without preference, i.e., uniformly at random. For triangulations, such elements are vertices, edges, and triangles. Then, (ii) a specific transformation from the set of operations that keep the simplicial complex intact is applied to this element. For triangulations, this transformation is one of the triangular mesh operations, in particular, operations $P1$, $P2$, $S$, and $S'$ described in Sec.~\ref{s4}. Depending on specific (i) and (ii), we get a wide range (zoo) of evolution scenarios, including, in general, growing, decaying, and equilibrium networks  
%%The set of these models forms a rich zoo of manifolds 
with diverse structures, space dimensions, and topologies which we describe in the following sections. 
We stress the following point. These transformations are interrelated, as we discussed in Sec.~\ref{s4}. Nonetheless, the progressive application of each specific transformation to preferentially selected elements of triangulation networks generates a distinct triangulation.  

For the sake of convenience and reference we introduce our models 
%%in this section 
and list them in Table~\ref{t1}. 
%%Note that 
%%in this list we focus on most representative models. 
This list contains 
%%models providing 
examples for a wide range of situations: equilibrium and growing simplicial complexes, triangulations and higher-dimensional simplicial complexes, and evolving triangulations homeomorphic to closed surfaces with varying in time topology [growing number of holes (genera)].  
We choose a set of specific models enabling us to demonstrate a wide spectrum of Hausdorff and spectral dimensions and degree distributions. 
The problem is that, typically, it is difficult to measure the Hausdorff and spectral dimensions in networks of sizes accessible for our simulations; see Sec.~\ref{s6}.   
For example, it is virtually impossible to distinguish $d_H$ equal, say, $20$ from $d_H=\infty$, valid for  small words, since it would demand extremely large, out of reach, network sizes. On the other hand, it is much easier to observe low $d_H$, say, $2$, $3$, or $4$ in networks of reasonable sizes. 
So, for demonstration purposes, we have to choose specific models and their parameters to enable us to observe a set of finite Hausdorff and spectral dimensions for network sizes accessible in our simulations. 

%%by For these situations, we choose particular models demonstrating 
%%We start with two simplest growing triangulations (for the full list of the considered rules, see Table~\ref{t1}). 
%%All but one of the algorithms 
%%listed in 
%%this table conserve the topology of a manifold during the evolution. 
Rules G1 and 
%%G1n 
G1d 
%are
 describe the same dynamics of the network geometry with flavor \cite{bianconi2015complex,bianconi2015b-complex,bianconi2016emergent,bianconi2016network}  for $s=-1$, $\beta=0$, and dimension respectively 
%%$d=3$ 
$3$ 
and 
%%$d=n+1$. 
$d+1$. Specifically the G1 rule is a 
%%generalization 
variation of the random Apollonian graph \cite{zhou2004random,zhang2006high,frieze2012certain} implemented for a two-simplicial complex triangulating a closed surface. 

Rule GW generates evolving topology, a two-simplicial complex triangulating an $h$-holed torus with a progressively growing number of holes $h$. %%These ``wormholes'' are created by the merging pairs of 
Each of these holes 
%%``wormholes''  
is created by the merging of a pair of 
uniformly randomly chosen triangles 
%%of edges 
in a triangulation network. 
%%The mergings play a role of a long-range shortcuts in a specific small-world network  \cite{watts1998collective}. 
Here the merging of two triangles means that each vertex and edge of one triangle joins the corresponding element of the second triangle forming a single triangle. 
According to our rule, when this happens, these two faces 
%%based on the merging triangles 
annihilate, leaving an empty space, namely, a hole 
%%``wormhole'' 
after them. 
We stress that this emergent empty space (triangle of edges without a face) does not belong to the simplicial complex, and so we show it by white color in the scheme for model GW in Table~\ref{t1}. 
The annihilation of the two merging faces guarantees that any edge in this complex has two adjacent triangular faces. 
This transformation changes the topology of the underlying closed surface and so, in contrast to the mesh operations in all of our other models, it cannot be reduced to the Pachner moves or operation $S$. 
Note that this rule forbids the merging of the nearest-neighboring and, also, second-nearest-neighboring  triangles, since this would produce edges lying outside of triangulations, double edges, or one-cycles.  
The mergings play a role of long-range shortcuts in a specific small-world network  \cite{watts1998collective}. 
Each merging of this kind reduces the Euler characteristic $\chi$ by $2$, 
resulting in $\chi= 2(1-h)$ and the first Betti number $b_1=2h$. 
During the evolution, each 
hole 
%%wormhole 
widens with time due to the first channel of the process GW; see Table~\ref{t1}.  
In model GW we do not consider the twisting of the merging triangles. The reason is that 
%%Note that twisting of the merging triangles is not essential in our problems, since 
any ``twisted'' configuration can be untwisted by a series of Pachner moves or by applying operation $S$. This point, however, deserves a more detailed discussion; see Sec.~\ref{s8}.  

%%Note also that in model E1, the evolution stacks if there is no nodes of degree $3$ in a network
We consider long-time asymptotics, which are independent of initial configurations for most of these models. Note, however, that in model E1, the evolution stacks if there are no vertices of degree $3$ in a network, and so initial configurations for this case have to contain such vertices. Note also that an initial graph in model GW has to be sufficiently large, since the evolution rule employs merging non-neighboring triangles. As is natural, all initial configurations for our models must be closed manifolds.

\begin{table*}
\caption{Zoo of complex network models of evolving closed manifolds} 
%%$t_c$ is the critical point, $\beta$ and $\tau$ are the critical exponents of the percolation cluster and the finite cluster size zistribution, respectively, $f(0)$ is the critical amplitude, and shows}
%%\begin{center}
%\begin{tabular}{c l c c}
\begin{tabular}{p{1cm} p{13.6cm} p{2.6cm} }
\hline
%\noalign{\smallskip}
%%& & & & 
%%\\
%%[-9pt]
{\footnotesize 
%%Notation 
Model 
}    
&
\hspace{3.4cm} 
{\footnotesize Operation at each step of evolution}     
&  
{\footnotesize \hspace{20pt}Scheme} 
\\[1pt]
\hline
%%\noalign{\smallskip}
%%\svhline
%%\noalign{\smallskip}
%%\hline
%%[0pt] 
%%& & & & 
\\[-11pt]
\multicolumn{3}{c}{Growing triangulations} 
\\[2pt]
\hline
\\[-15pt]
%%%%%\hline
%%%%%&&&&
%%%%%\\
G1   
&
\parbox[c]{12cm}{
%%{\scriptsize 
{\footnotesize 
\flushleft
(i) choose a triangle uniformly at random,
\\
(ii) attach a new vertex to all three vertices of this triangle (Pachner's 0-move). 
\\
{\scriptsize 
(This rule is closely related to the one governing the evolution of random Apollonian networks \cite{zhou2004random,zhang2006high,frieze2012certain}. The difference is that the manifold is closed here.) 
}
\\
}
}
&  
\parbox[c]{2cm}{\centering \includegraphics[width=2.6cm]{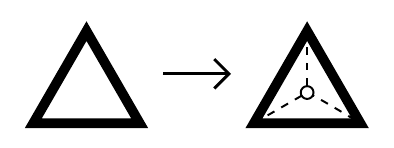}}
\\
G2  
& 
\parbox[c]{12cm}{
%%{\scriptsize 
{\footnotesize 
\flushleft
(i) choose an edge uniformly at random,
\\
(ii) exchange it for a new vertex attached to all four vertices of the two triangles sharing this edge.
\\
} 
}
& 
\parbox[c]{2cm}{\centering \includegraphics[width=2.6cm]{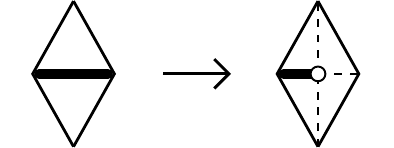}}
\\
G  
& 
\parbox[c]{12cm}{
%%{\scriptsize 
{\footnotesize 
\flushleft
(i) choose a vertex uniformly at random and two its random edges,
\\
(ii) split them in the way shown in Fig.~\ref{f3}, the move from left to right.
\\
}
}
& 
\parbox[c]{2cm}{\centering \includegraphics[width=2.6cm]{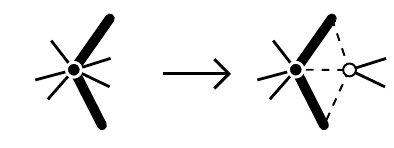}}
\\
Ga 
& 
\parbox[c]{12cm}{
%%{\scriptsize 
{\footnotesize 
\flushleft
(i) choose a vertex uniformly at random and one of its edges at random;
%%\\
then, among the rest edges of the vertex, if the vertex degree is even, choose the opposite edge to the first, 
%%\\
if the degree is odd, choose, with equal probability, one of the two most remote edges (here remote, relatively to the first); 
\\
(ii) split them in the way shown in Fig.~\ref{f3}, the move from left to right.
\\
}
}
& 
\parbox[c]{2cm}{\centering \includegraphics[width=2.6cm]{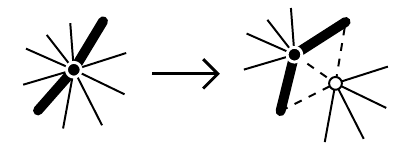}}
\\
Gb  
&
\parbox[c]{12cm}{
%%{\scriptsize 
{\footnotesize 
\flushleft
(i) 
choose a vertex uniformly at random and one of its edges at random; then, among the rest edges of the vertex, choose, with equal probability, one of the two closest (to the first) edges;
\\
(ii) split them in the way shown in Fig.~\ref{f3}, the move from left to right.
\\
{\scriptsize 
(Gb is close to rule G1. The difference is that here a random triangle incident to a uniformly chosen vertex is selected.)
}
\\
}
}
& 
\parbox[c]{2cm}{\centering \includegraphics[width=2.6cm]{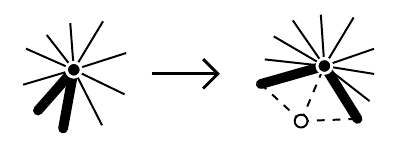}}
\\
Gc  
& 
\parbox[c]{12cm}{
%%{\scriptsize 
{\footnotesize 
\flushleft
(i)
choose an edge uniformly at random and its end vertex with the highest number of connections 
\\
(if the degrees of the ends coincide, then choose any one of them with equal probability), 
\\
(ii) choose the second edge as in rule Ga,  
\\
(iii) split the two chosen edges and the vertex in the way shown in Fig.~\ref{f3}, the move from left to right.
\\
}
}
& 
\parbox[c]{2cm}{\centering \includegraphics[width=2.6cm]{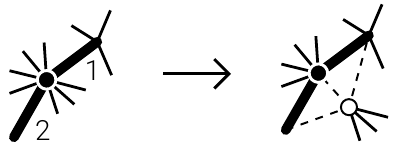}}
\\
G$'$  
& 
\parbox[c]{12cm}{
%%{\scriptsize 
{\footnotesize 
\flushleft
(i) choose a vertex uniformly at random and two its random edges except those belonging to the same triangle,  
\\
(ii) split them in the way shown in Fig.~\ref{f5}, the move from left to right.
\\
}
}
& 
\parbox[c]{2cm}{\centering \includegraphics[width=2.6cm]{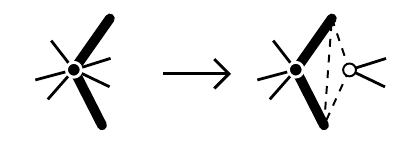}}
\\[-5pt]
\\
\hline
\multicolumn{3}{c}{Growing 
%%$n$-dimensional 
$d$-dimensional 
%%manifolds} 
simplicial complexes}
\\[2pt]
\hline
\\[-15pt]
%%G1n   
G1d 
& 
\parbox[c]{15cm}{
%%{\scriptsize 
{\footnotesize 
\flushleft
(i)
choose a simplicial complex uniformly at random, and 
\\
(ii)
attach a new vertex to all 
%%$n+1$ 
$d+1$ vertices of this simplicial complex. 
\\
{\scriptsize
(Note that 
%%G1n 
G1d directly generalizes G1 to an arbitrary $d\geq2$.) 
%%$n\geq2$.) 
}
\\
}  
}
&																						
\parbox[c]{2cm}{\centering \includegraphics[width=2.6cm]{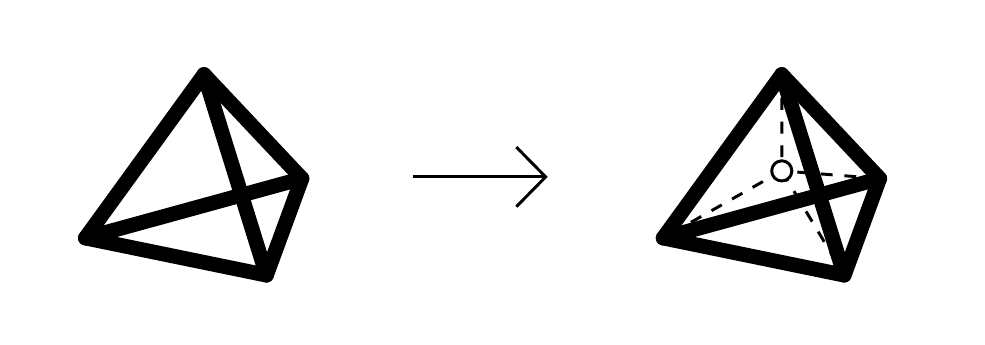}}
\\
%%G2n  
G2d
& 
\parbox[c]{12cm}{
%%{\scriptsize 
{\footnotesize 
\flushleft
(i) 
choose a $(d{-}1)$-simplex 
%%an $(n{-}1)$-simplex 
uniformly at random,   
\\
(ii) 
attach a new vertex to all $d+2$ 
%%$n+2$ 
vertices of the two $d$-simplices 
%%$n$-simplices 
sharing the chosen $(d{-}1)$-simplex. 
%%$(n{-}1)$-simplex.
\\
{\scriptsize
(Note that 
%%G2n 
G2d is defined only for $d\geq3$, 
%%$n\geq3$, 
and so G2d 
%%G2n 
does not generalize G2 directly.) 
}
\\
}
}
& 
\parbox[c]{2cm}{\centering \includegraphics[width=2.6cm]{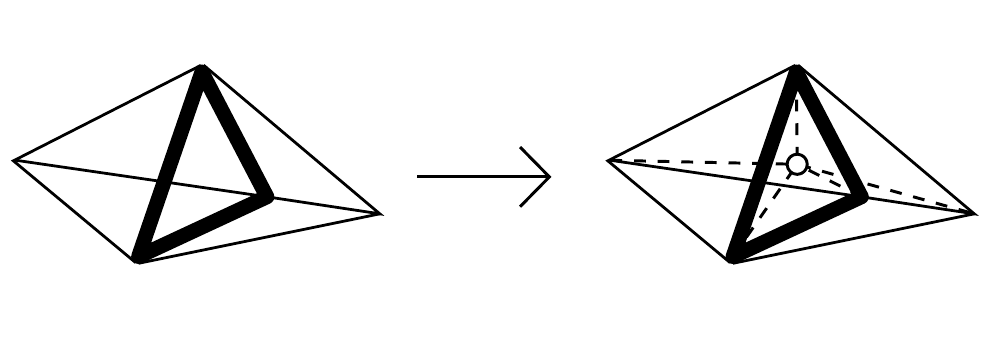}}
%%\\
\\[27pt]
\hline
\\[-13pt]
\multicolumn{3}{c}{Generation of holes in a growing triangulation} 
\\[2pt]
\hline
       \\[-15pt]
       GW   
       & 
        \parbox[c]{12cm}{
%%%%%%%{\scriptsize 
        {\footnotesize 
         \flushleft
         (i) 
at each step perform rule G2, and, in addition, 
          \\
(ii) at each $\theta$-th step choose two 
%%non-neighboring (and non-second-neighboring) 
triangles, excluding first- and second-neighboring ones, uniformly at random and merge them into a single triangle. These two merging faces annihilate creating a hole (genus) in the triangulation and in the corresponding surface. 
%%%%%\\
          \\
           }  
           }
          &  
          $\phantom{}$\hspace{-18pt}\parbox[c]{3.2cm}{\vspace{8pt}\centering \includegraphics[width=3.2cm]{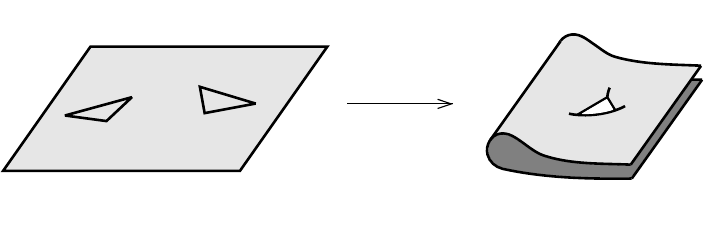}}  
            \\
            \\[-5pt]
\hline
\\[-11pt]
\multicolumn{3}{c}{Equilibrium triangulations} 
\\[2pt]
\hline
\\[-15pt]
E1   
& 
\parbox[c]{12cm}{
%%{\scriptsize 
{\footnotesize 
\flushleft
(i) choose a vertex of degree $3$ uniformly at random and remove it, 
\\
(ii) choose a triangle uniformly at random and attach a new vertex to all three vertices of this triangle.
\\
}  
}
&
\parbox[c]{2cm}{\centering \includegraphics[width=2.6cm]{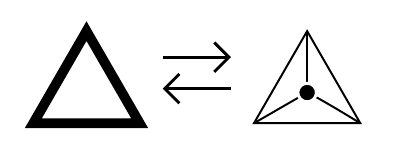}}
\\
E2  
& 
\parbox[c]{12cm}{
%%{\scriptsize 
{\footnotesize 
\flushleft
(i) choose an edge uniformly at random,
\\
(ii) perform Pachner 1-move (flip) with this edge (see Fig.~\ref{f1}, P2).
\\
}
}
& 
\parbox[c]{2cm}{\centering \includegraphics[width=2.6cm]{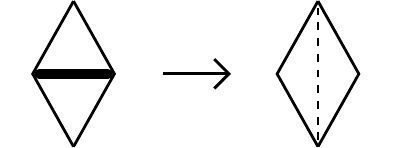}}
\\
E3  
& 
\parbox[c]{12cm}{
%%{\scriptsize 
{\footnotesize 
\flushleft
(i)
choose an edge uniformly at random and compress it into one vertex as in transformation S, Fig.~\ref{f3}, the move from left to right,
\\
(ii) make a step according to rule G.
\\
}  
}
& 
\parbox[c]{2cm}{\centering \includegraphics[width=2.6cm]{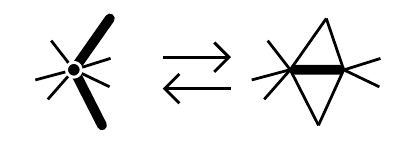}}
\\
\\[-9pt]

\noalign{\smallskip}\hline\noalign{\smallskip}
%%[2pt]
%%\hline
\end{tabular}
%%\end{center}
\label{t1}
\end{table*}

%%%%%%%%%%%%%%%%%%%%%%%%%%%%%%%%%%%%%%%%%%%%%%
%%%%%%%%%%%%%%%%%%%%%%%%%%%%%%%%%%%%%%%%%%%%%%
%%%%%%%%%%%%%%%%%%%%%%%%%%%%%%%%%%%%%%%%%%%%%%
%%%%%%%%%%%%%%%%%%%%%%%%%%%%%%%%%%%%%%%%%%%%%%

%%
%%

%%%%%%%%%%%%%%%
%%%%%%%%%%
%%%%%%%%%%%%%%
%%%%%%%%%%%

\section{Local properties}
\label{s5}

\begin{figure}[t]
\begin{center}
\includegraphics[scale=0.52]{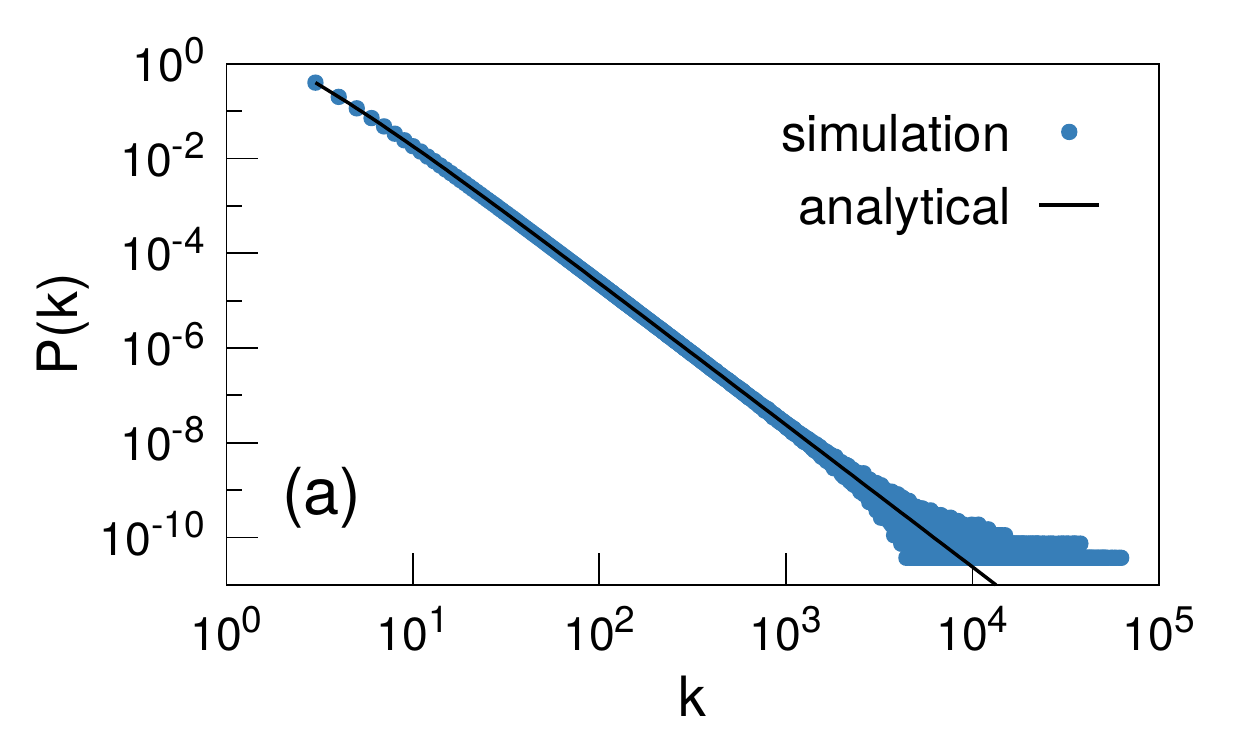}
\\[12pt]
\includegraphics[scale=0.52]{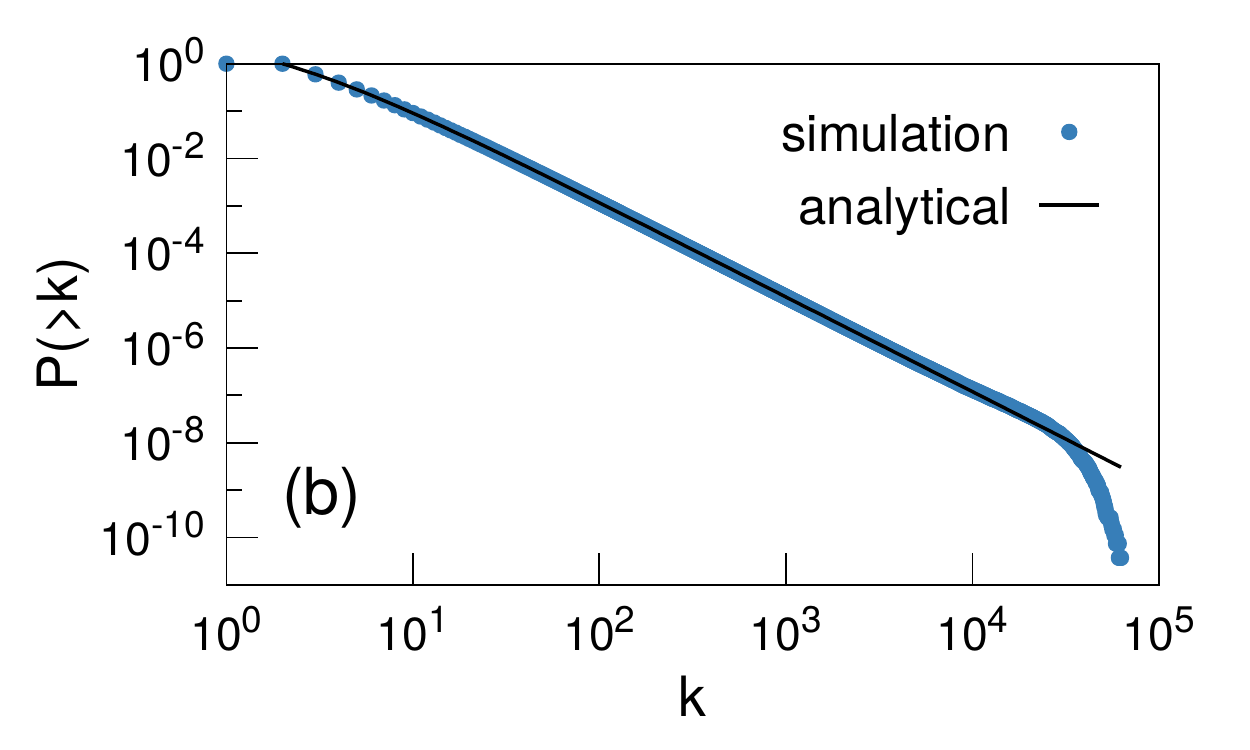}
\end{center}
\caption{
(a) The degree distribution of model G1: the results of numerical simulations (network of $2^{27}$ vertices, 200 samples) and the theoretical curve, Eq.~(\ref{80}). (b) The corresponding cumulative degree distribution $P_\text{cum}(k)=\sum_{q\geq k}P(q)$, simulations 
%%(network of $2^{27}$ vertices) 
and the expression $12/[k(k+1)]$, obtained from Eq.~(\ref{80}). 
%%The numerical simulations employ 200 samples. 
}
\label{f8}       
\end{figure}

\begin{figure}[t]
\begin{center}
\includegraphics[scale=0.52]{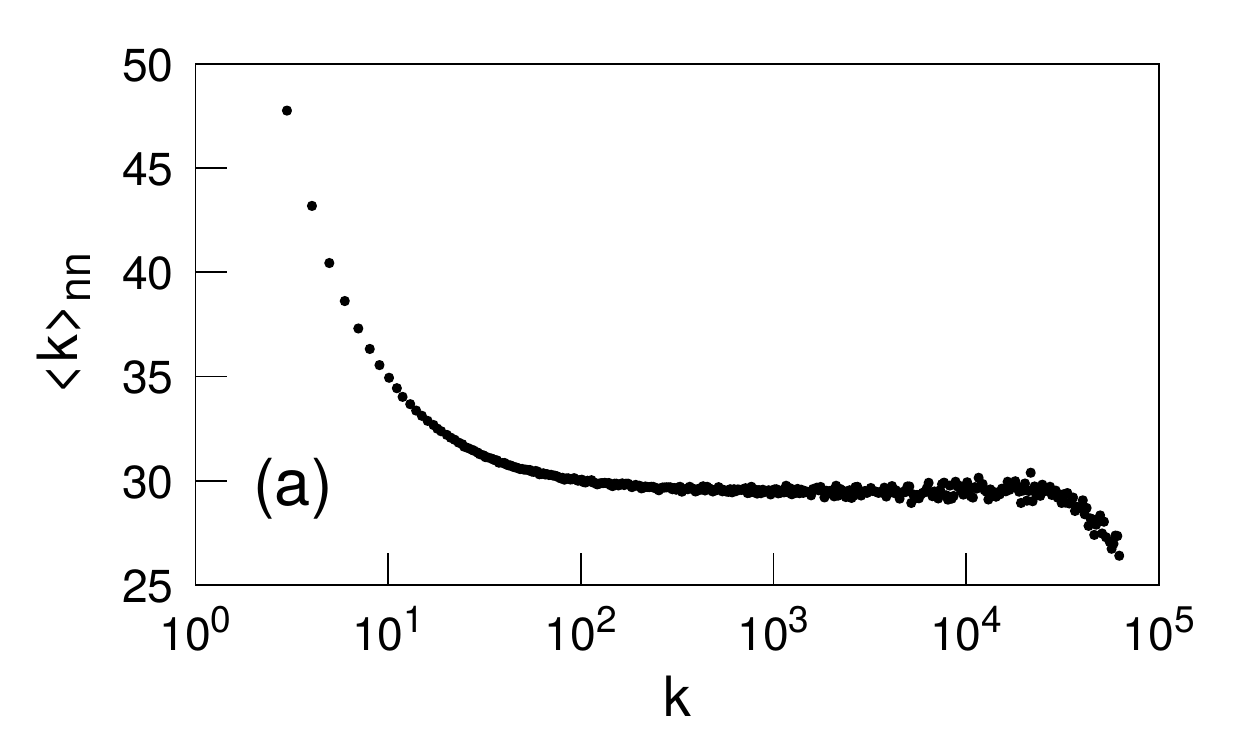}
\\[12pt]
\includegraphics[scale=0.52]{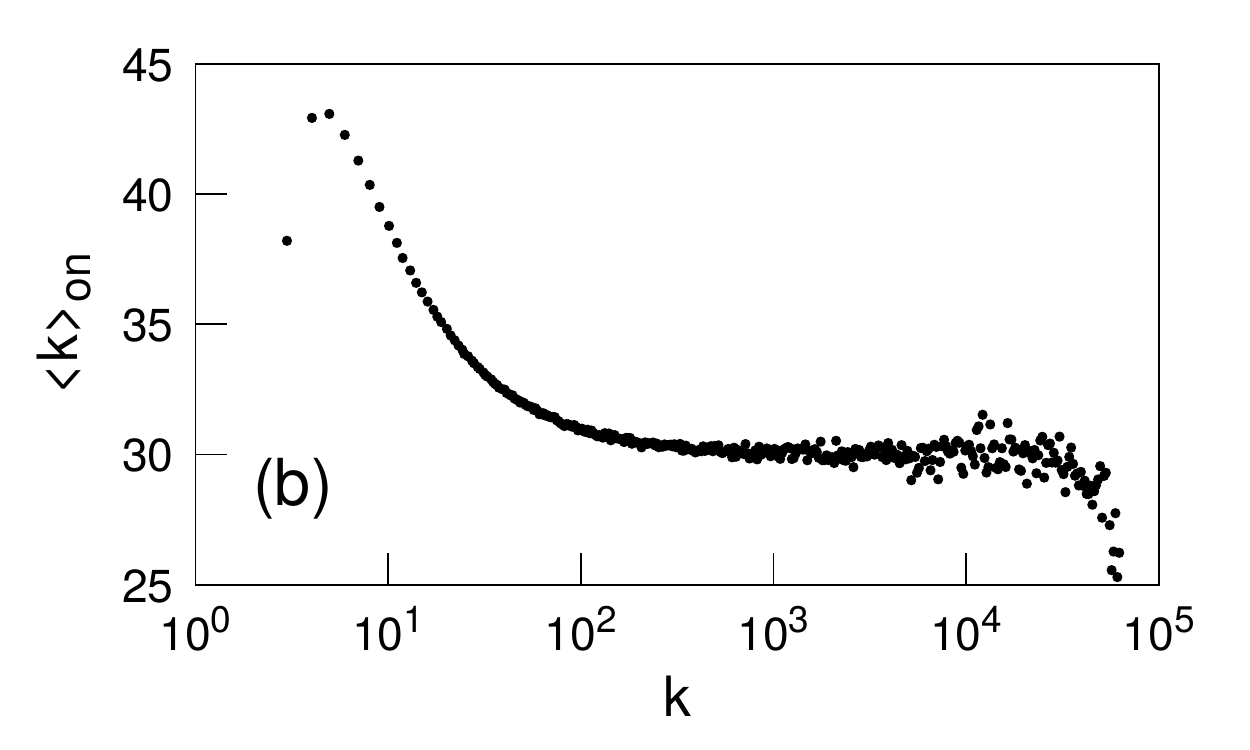}
\end{center}
\caption{ 
Degree--degree correlations in model G1. 
(a) The average degree $\langle k \rangle_ \text{nn}(k)$ of the nearest neighbors of a vertex of degree $k$. 
 (b) The average degree $\langle k \rangle_ \text{on}(k)$ of the vertex in a rhombus, opposite to a vertex of degree $k$, that is not its nearest neighbor.  The points in the plots are obtained by logarithmic binning the results of numerical simulations for a network of $2^{27}$ vertices. The numerical simulations employ 200 samples. 
}
\label{f9}       
\end{figure}

We start considering the local properties of these complex networks with their degree distributions, $P(k)$. 
%%Analytical expressions can be easily derived if we neglect degree--degree correlations. 
The degree distributions for the models G1 and G2 of growing triangulations can be derived analytically. 
For the G1 model for growing triangulations, we write the following rate equation describing the evolution of the average number $N(k,t)=tP(k,t)$ of vertices of degree $k$ in the network at time $t$ (current number of steps): 
\begin{eqnarray}
N(k,t&+&1) = N(k,t) 
\nonumber
\\[3pt]
&+& 3 \frac{k-1}{\langle k \rangle t}N(k-1,t) - 3 \frac{k}{\langle k \rangle t}N(k,t) + \delta_ {k,3}  
, 
\label{60}
\end{eqnarray}
which directly generalizes the equations for a network growing by attachment of a new vertex to the ends of a randomly chosen edge \cite{dorogovtsev2001size}. Here the average degree 
%%$\langle k \rangle$ 
of a node,  $\langle k \rangle(t)$, approaches 6 as $t\to\infty$. The form of the right-hand side of this equation directly follows from the fact that the degree distribution of the vertices of a randomly chosen triangle of a triangulation is $kP(k)/\langle k \rangle$. Note that this is the case, independently of correlations in the network. In the infinite size, the degree distribution is stationary and independent of initial configuration. (Recall that in this work we consider only closed manifolds, so an initial configuration is also closed.) 
In this limit, Eq.~(\ref{60}) is reduced to 
\begin{equation}
P(k) = \frac{1}{2}(k-1)P(k-1) -  \frac{1}{2}kP(k) + \delta_ {k,3} 
.
\label{70}
\end{equation}
The solution of this equation is the degree distribution 
\begin{equation}
P(k\geq3) =\frac{24}{k(k+1)(k+2)}
, 
\label{80}
\end{equation}
%%
%%which is exactly the same as for 
which is in total agreement with the result obtained in \cite{bianconi2016network,bianconi2015complex}.
Interestingly also if the network has large clustering coefficient, the obtained degree distribution is exactly the same of the standard Barab\'asi--Albert model \cite{barabasi1999emergence,dorogovtsev2000structure} in the particular case of attachment of a new vertex to three existing ones. The resulting degree distribution exponent $\gamma$ equals $3$. 
%%and directly generalizes the equations for a network growing by attachment of a new vertex to the ends of a randomly  
The degree distribution plot, Fig.~\ref{f8}(a), and the cumulative distribution $P_\text{cum}(k)=\sum_{q=k}^\infty P(q)$, Fig.~\ref{f8}(b), demonstrate that the analytical result completely agrees with numerical simulations. Figure~\ref{f9}(a) showing the average degree $\langle k \rangle_\text{nn}(k)$ of the nearest neighbors of a vertex of degree $k$ indicates that the degrees of nearest neighbors in this network are correlated but these correlations are not strong similarly to the Barab\'asi--Albert model (see the region of large $k$). It is natural to consider another type of degree--degree correlations in triangulation networks.  Choose an edge uniformly at random. Two triangles share this edge. We consider correlations between the degrees of the two vertices of these triangles that are not ends of this joint edge. 
%%, and so they belong to different triangles. 
These vertices are second nearest neighbors of each other. The point is that they are the closest vertices that belong to different triangles (faces) in this network. 
In other words, these are two opposite vertices of a rhombus, and there is no edge between them. 
We characterize these correlations by $\langle k \rangle_\text{on}(k)$ which is the average degree of such neighbors of a vertex of degree $k$. Figure~\ref{f9}(b) demonstrates these degree--degree correlations in network G1. Notice that, apart from the region of low $k$, these two curves are qualitatively similar to each other despite the stronger separation of vertices in the second case. 

\begin{figure}[t]
\begin{center}
\includegraphics[scale=0.52]{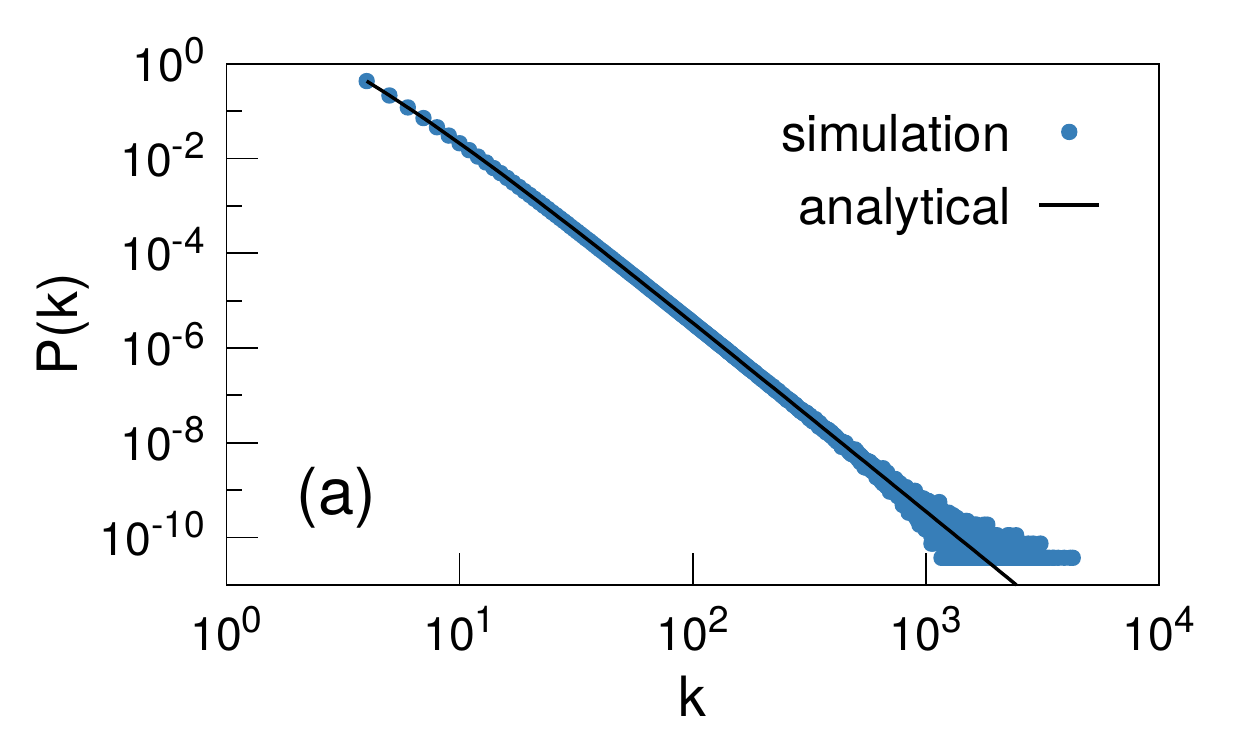}
\\[12pt]
\includegraphics[scale=0.52]{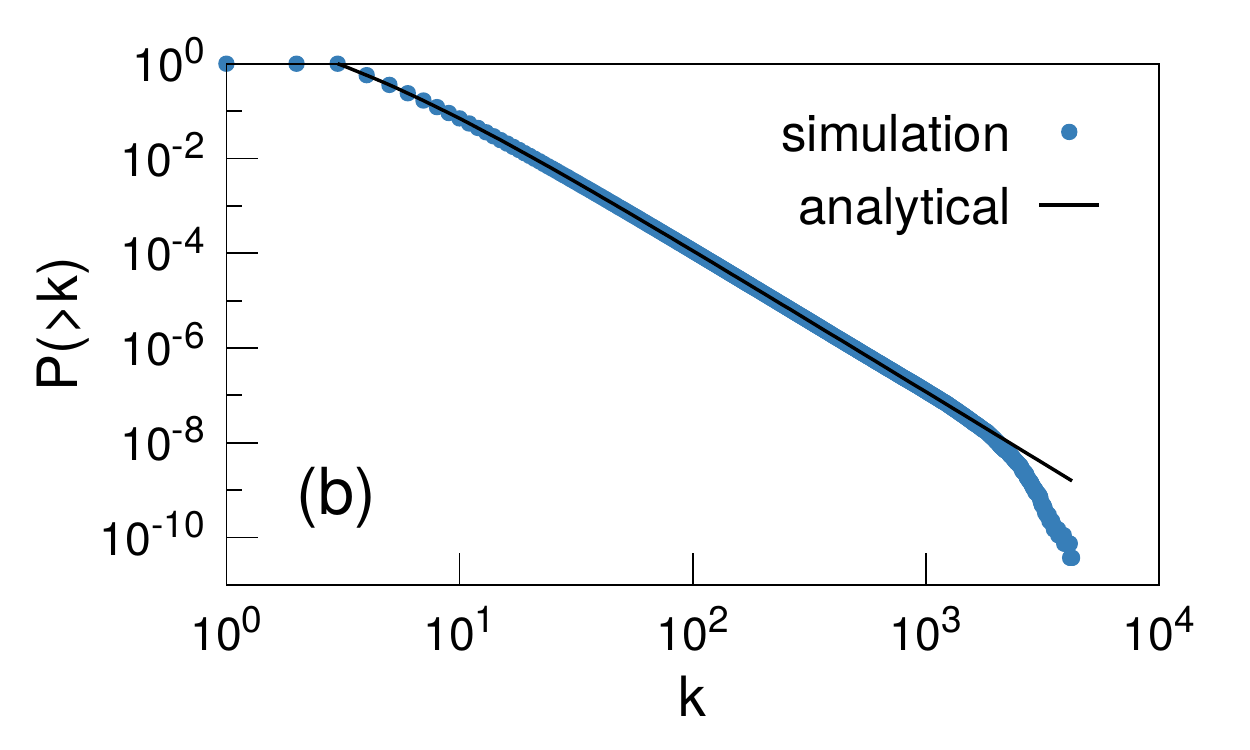}
\end{center}
\caption{
(a) The degree distribution of model G2: the results of numerical simulations (network of $2^{27}$ vertices, 200 samples) and the theoretical curve, Eq.~(\ref{110}). (b) The corresponding cumulative degree distribution $P_\text{cum}(k)$, simulations, and the expression $120/[k(k+1)(k+2)]$, obtained from Eq.~(\ref{110}).  
%%(network of $2^{27}$ nodes) The numerical simulations employ 200 samples. 
}
\label{f10}       
\end{figure}

\begin{figure}[t]
\begin{center}
\includegraphics[scale=0.52]{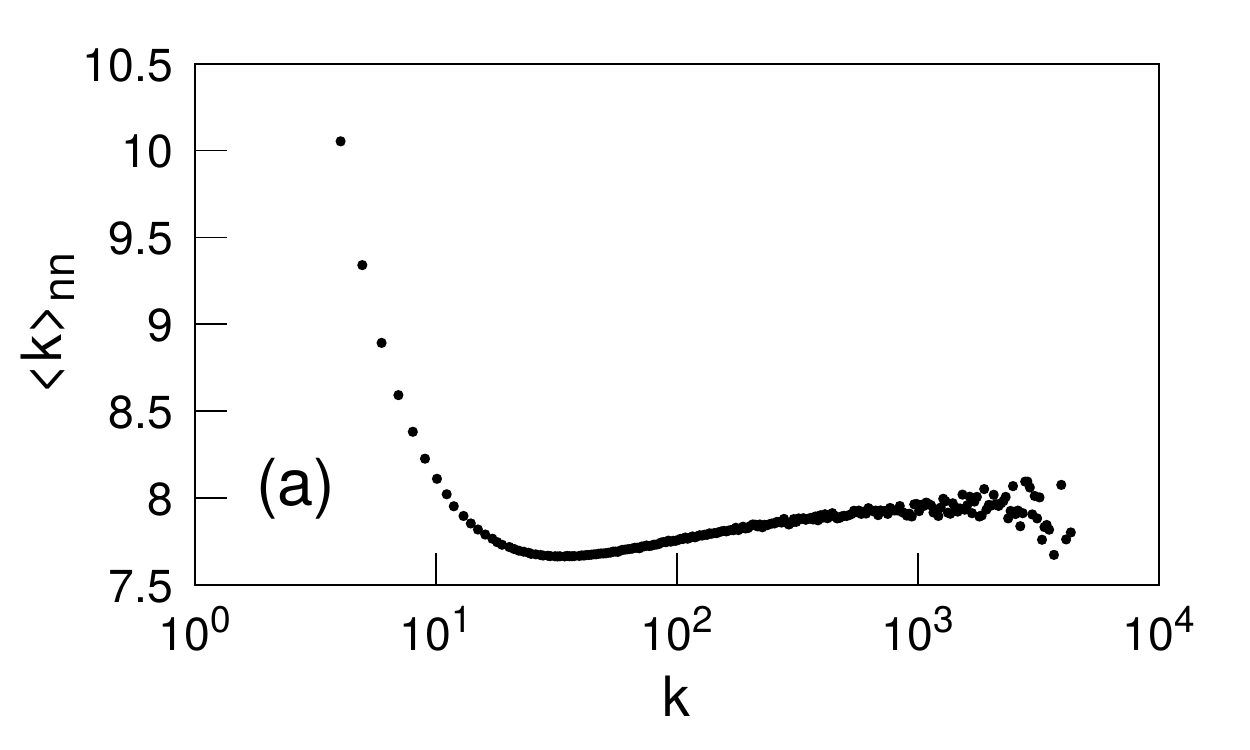}
\\[12pt]
\includegraphics[scale=0.52]{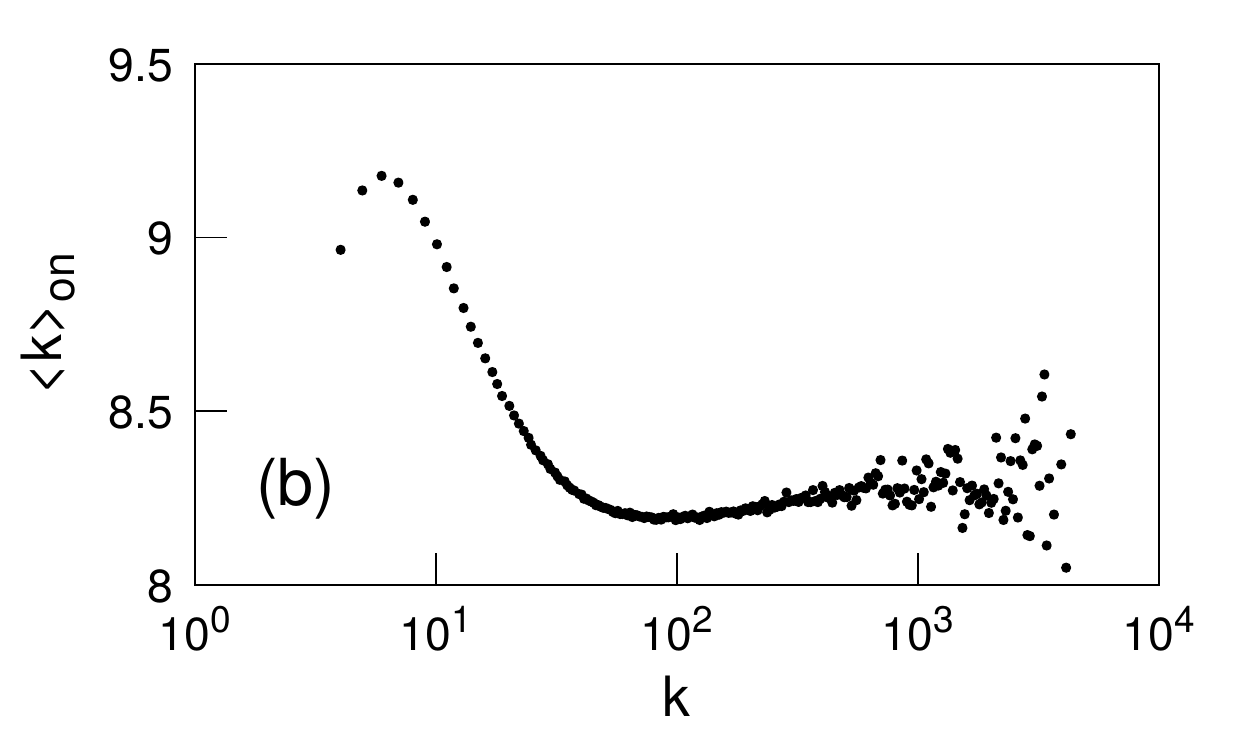}
\end{center}
\caption{
Degree--degree correlations in model G2. 
The average degree $\langle k \rangle_ \text{nn}(k)$ of the nearest neighbors of a vertex of degree $k$. 
 (b) The average degree $\langle k \rangle_ \text{on}(k)$ of the vertex in a rhombus, opposite to a vertex of degree $k$, that is not its nearest neighbor.  The points in the plots are obtained by logarithmic binning the results of numerical simulations for a network of $2^{27}$ vertices. The numerical simulations employ 200 samples. 
%%(a) DEGREE-DEGREE CORRELATIONS IN G2 FOR NN  and (b) DEGREE-DEGREE CORRELATIONS IN G2 FOR the opposite vertices of two adjacent triangles (SECOND NN). (network of $2^{27}$ nodes) The numerical simulations employ 200 samples. 
}
\label{f11}       
\end{figure}

The model G2 turns out to be more interesting. Its structure strongly deviates from G1 and the Barab\'asi--Albert model. The rate equations for triangulations growing according to rule G2 have the following form: 
\begin{eqnarray}
N(k,t&+&1) = N(k,t) 
\nonumber
\\[3pt]
&+& 2 \frac{k-1}{\langle k \rangle t}N(k-1,t) - 2 \frac{k}{\langle k \rangle t}N(k,t) + \delta_ {k,4}  
\label{90}
\end{eqnarray}
with, similarly to Eq.~(\ref{60}), the mean degree $\langle k \rangle(t)$ approaching 6 as $t\to\infty$. In this limit, we get 
\begin{equation}
P(k) = \frac{1}{3}(k-1)P(k-1) -  \frac{1}{3}kP(k) + \delta_ {k,4} 
, 
\label{100}
\end{equation}
and so the degree distribution is 
\begin{equation}
P(k\geq4) =
%%\frac{6!}{2}\frac{(k-1)!}{(k+3)!}
\frac{360}{k(k+1)(k+2)(k+3)}
,  
\label{110}
\end{equation}
which provides the asymptotics: $P(k) \sim k^{-4}$. Consequently, the degree distribution exponent $\gamma$ equals $4$ for this network. Figure~\ref{f10} validates the analytical results by comparison with numerical simulations. In contrast to G1, this network is essentially correlated as one can see from Fig.~\ref{f11}(a) showing the average degree $\langle k \rangle_\text{nn}(k)$ of the nearest neighbors of a vertex of degree $k$ in network G2. Notice that these correlations are assortative in the region of large degrees. This is similar to recursive preferentially growing scale-free networks with $\gamma>3$. 
Figure~\ref{f11}(b) shows the plot of $\langle k \rangle_\text{on}(k)$ for this network. Notice that the second plot indicates at least not weaker degree-degree correlations than in Fig.~\ref{f11}(a), despite  the stronger separation of the ``on'' vertices. Note also that in model G2, the ``on'' neighbors enter in the evolution rule, which justifies the consideration of these correlations.

%%%%%%%%%%%%%%%%%%%%%%%%%%%%%%%%%%%%%%%%%%%%%%
%%%%%%%%%%%%%%%%%%%%%%%%%%%%%%%%%%%%%%%%%%%%%%

%%
%%

%%%%%%%%%%%%%%%%%%%%%%%%%%%%%%%%%%%%%%%%%%%%%%
%%%%%%%%%%%%%%%%%%%%%%%%%%%%%%%%%%%%%%%%%%%%%%

%%%%%%%%%%%%%%%%%%%%%%%%%%%%%%%%%%%%%%%%%%%%%%
%%%%%%%%%%%%%%%%%%%%%%%%%%%%%%%%%%%%%%%%%%%%%%

%%
%%

%%%%%%%%%%%%%%%%%%%%%%%%%%%%%%%%%%%%%%%%%%%%%%
%%%%%%%%%%%%%%%%%%%%%%%%%%%%%%%%%%%%%%%%%%%%%%
    
%%%%%%%%%%%%%%%%%%%%%%%%%%%%%%%%%%%%%%%%%%%%%%
%%%%%%%%%%%%%%%%%%%%%%%%%%%%%%%%%%%%%%%%%%%%%%

%%
%%

%%%%%%%%%%%%%%%%%%%%%%%%%%%%%%%%%%%%%%%%%%%%%%
%%%%%%%%%%%%%%%%%%%%%%%%%%%%%%%%%%%%%%%%%%%%%%
    
%%%%%%%%%%%%%%%%%%%%%%%%%%%%%%%%%%%%%%%%%%%%%%
%%%%%%%%%%%%%%%%%%%%%%%%%%%%%%%%%%%%%%%%%%%%%%

%%
%%

%%%%%%%%%%%%%%%%%%%%%%%%%%%%%%%%%%%%%%%%%%%%%%
%%%%%%%%%%%%%%%%%%%%%%%%%%%%%%%%%%%%%%%%%%%%%%

We also considered a more complicated mixed model in which at each step rule G1 is realized with probability $p$ while rule G2 is realized with the complementary probability $1-p$. A straightforward calculation resulted in a power-law degree distribution with the exponent 
\begin{equation}
\gamma = 3 + \frac{2(1-p)}{2+p}
.  
\label{115}
\end{equation}

The asymptotics of the degree distributions for rules G1 and G2 can be also found for higher-dimensional manifolds. We use the following well-known expression for the degree distribution exponent of scale-free networks growing due to preferential attachment. If the preference function for the probability of attachment to a vertex of degree $k$  is $k+A$, where $A$ is additional attractiveness, and each new vertex has $m$ edges, then $\gamma=3+A/m$  \cite{dorogovtsev2000structure}. 

In model G1d, 
%%G1n,   
%%, Fig.~\ref{f6}, 
each vertex attains new connections with probability proportional to the number of simplexes incident to this vertex. This number, $s=s(k,d)$, 
%%$s=s(k,n)$, 
is expressed in terms of the degree of a vertex $k$ and the dimensionality $d$ 
%%$n$ 
of a simplicial complex in the following way: 
\begin{equation}
s(k,d) = d+1 + [k-(d+1)](d-1)
%%s(k,n) = n+1 + [k-(n+1)](n-1)
,   
\label{120}
\end{equation}
where $k\geq d+1$. 
%%$k\geq n+1$. 
[Notice that, in particular, $s(k,2)=k$ for triangulations, and $s(d+1,d)=d+1$, 
%%$s(n+1,n)=n+1$, 
as is natural.]  
Consequently, the preference function in model G1d 
%%G1n 
is $k-(d{+}1)+(d{+}1)/(d{-}1)$, 
%%$k-(n{+}1)+(n{+}1)/(n{-}1)$, 
which leads to  
\begin{equation} 
\gamma = 2 + \frac{1}{d-1}
%%\gamma = 2 + \frac{1}{n-1}
%%.  
\label{130}
\end{equation}
since here the number of edges of a new vertex is $m=d+1$. 
%%$m=n+1$. 
For triangulations ($d=2$) 
%%($n=2$) 
this confirms the degree distribution exponent $\gamma=3$ in model G1; see Eq.~(\ref{80}). According to Eq.~(\ref{130}) exponent $\gamma$ for model G1d 
%%G1n 
is in the range between  $2$ ($d\to\infty$) 
%%($n\to\infty$) 
and $3$ ($d=2$). 
%%($n=2$). 
These results therefore confirm the results obtained in Refs.~\cite{bianconi2016network,bianconi2015complex,bianconi2015b-complex} for manifolds of dimension %%$d=n+1$ 
$d+1$ with a boundary.

In contrast to G1d, 
%%G1n, 
in our model G2d 
%%G2n 
for $(d{>}2)$-dimensional 
%%$(n{>}2)$-dimensional 
growing simplicial complexes, there are actually two kinds of attachment. Of $d+2$ 
%%$n+2$ 
edges of a new vertex, $d$ 
%%$n$ 
are attached to the vertices of a randomly chosen $(d{-}1)$-simplex 
%%$(n{-}1)$-simplex 
and the remaining two are attached to the two vertices of the two $d$-simplices 
%%$n$-simplices 
sharing the $(d{-}1)$-simplex. 
%%$(n{-}1)$-simplex. 
At first sight, these two channels principally differ from each other. This is, however, not the case. Let us consider the number $s_{d-1}=s_{d-1}(k,d)$ 
%%$s_{n-1}=s_{n-1}(k,n)$ 
of $(d-1)$-simplices 
%%$(n-1)$-simplices 
incident to a vertex (of degree $k$) in a $d$-dimensional 
%%$n$-dimensional 
simplicial complex and find how this number is related to the number $s=s(k,d)$ 
%%$s=s(k,n)$ 
of $d$-simplices 
%%$n$-simplices 
incident to this vertex, Eq.~(\ref{120}). 
Each $d$-simplex 
%%$n$-simplex 
has $d+1$ 
%%$n+1$ 
faces [$(d{-}1)$-simplices], 
%%[$(n{-}1)$-simplices], 
and each of its vertices is on $d$ 
%%$n$ 
of those. Then the number of these faces incident to a vertex is  
\begin{equation}
s_{d-1}(k,d) = \frac{d}{2}s(k,d)
%%s_{n-1}(k,n) = \frac{n}{2}s(k,n)
.  
\label{140}
\end{equation}
Here the factor $1/2$ is due to the fact that each face is shared by a pair of $d$-simplices. 
%%$n$-simplices. 
[Note that when $d$ 
%%$n$ 
is odd, the number $s(k,d)$ 
%%$s(k,n)$ 
is even, see Eq.~(\ref{120}), which guarantees that $s_{d-1}(k,d)$ 
%%$s_{n-1}(k,n)$ 
is an integer]. Thus, since $s_{d-1}(k,d)$ 
%%$s_{n-1}(k,n)$ 
is proportional to $s(k,d)$, 
%%$s(k,n)$, 
then the preference function for both channels of attachment is the same, $k-(d{+}1)+(d{+}1)/(d{-}1)$, 
%%$k-(n{+}1)+(n{+}1)/(n{-}1)$, 
as in the G1 model. Now, however, we have $d+2$ 
%%$n+2$ 
attachments of  a new vertex, so we get the degree distribution exponent 
\begin{equation} 
\gamma = 2 + \frac{2d}{(d-1)(d+2)}
%%\gamma = 2 + \frac{2n}{(n-1)(n+2)}
%%.  
\label{150}
\end{equation}
for $d\geq 3$. 
%%$n\geq 3$. 

We considered in detail only the power-law distributions. These distributions were generated by models of growing closed manifolds G1 and G2 (growing triangulations) and G1d 
%%G1n 
and G2d. 
%%G2n. 
For the other models of growing triangulations discussed in this paper, namely models G, Ga, Gb, Gc, and G$'$, our simulations showed less interesting degree distributions decaying faster than a power law. For our models E1 and E2 of equilibrium triangulations, we reached network sizes of $2^{17}$, which turned out to be not sufficient to arrive at a reliable conclusion about the asymptotic form of the degree distributions, though we observed a faster decay than a power law. Simulations of network E3 of this size provided a power-law degree distribution with exponent $\gamma$ close to $3$.  
We list the resulting degree distribution exponents for our models in Table~\ref{t2}, where we indicate $\infty$ for $\gamma$ if the corresponding degree distribution decays more rapidly than a power-law function.

%%%%%%%%%%%%%%%%%%%%%%%%%%%%%%
%%%%%%%%%%%%%%%%%%%%%%%%%%%%%%
%%%%%%%%%%%%%%%%%%%%%%%%%%%%%%
%%%%%%%%%%%%%%%%%%%%%%%%%%%%%%

\section{Space dimensions}
%%{The method}
\label{s6} 

The key characteristic of the metric structure of a network is its space dimension. For the class of networks considered in this work it coincides with the Hausdorff dimension, so we denote it by $d_H$. For small worlds, i.e., networks whose diameter increases with size (number of vertices in a network, $N$) slower than a power-law function, $d_H=\infty$. There are two main methods to obtain the space dimension: (i) by measuring the asymptotic dependence of the average shortest path distance $\langle\ell \rangle$ between two uniformly randomly chosen vertices on the network size, $\langle\ell \rangle\propto N^{1/d_H}$; (ii) by measuring the asymptotic dependence of the number of vertices $V(r)$ within a sphere around a uniformly randomly chosen vertex in a large network on the radius $r$ of this sphere, $V(r) \propto r^{d_H}$. We mostly use the second method as it is more practical. 

\begin{figure}[t]
\begin{center}
\includegraphics[scale=0.52]{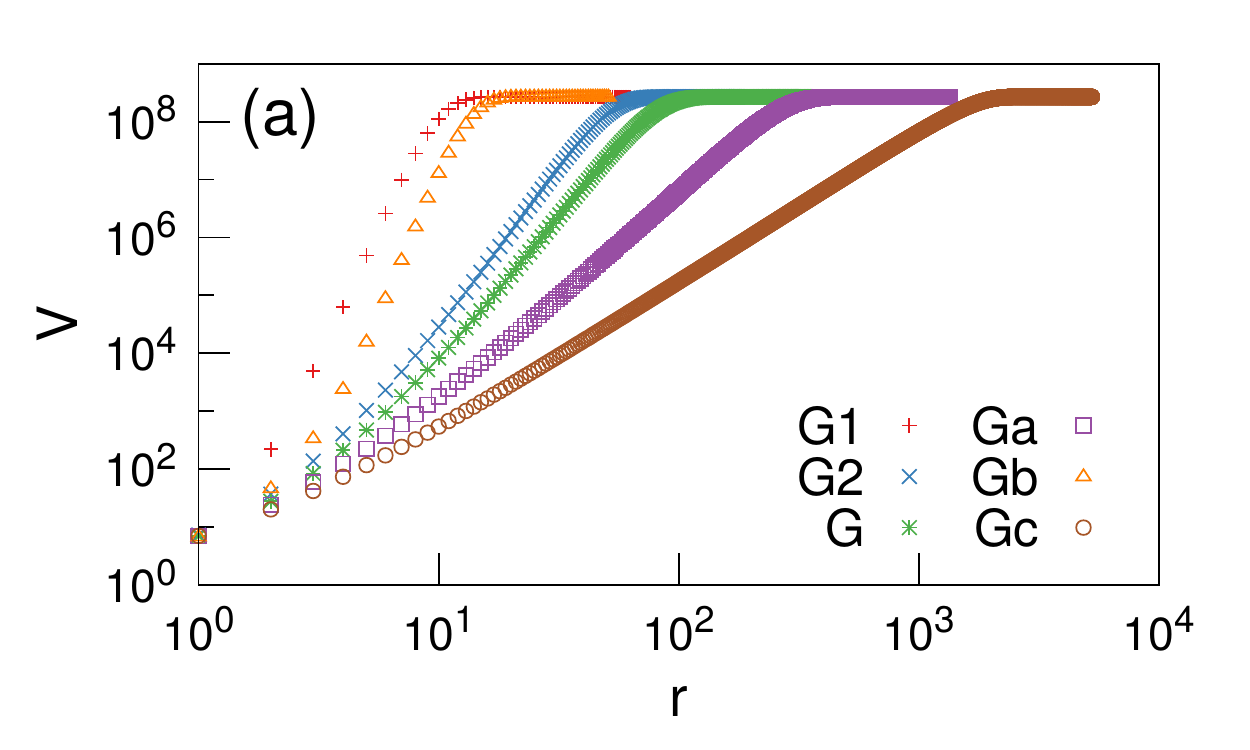}
\\[12pt]
\includegraphics[scale=0.52]{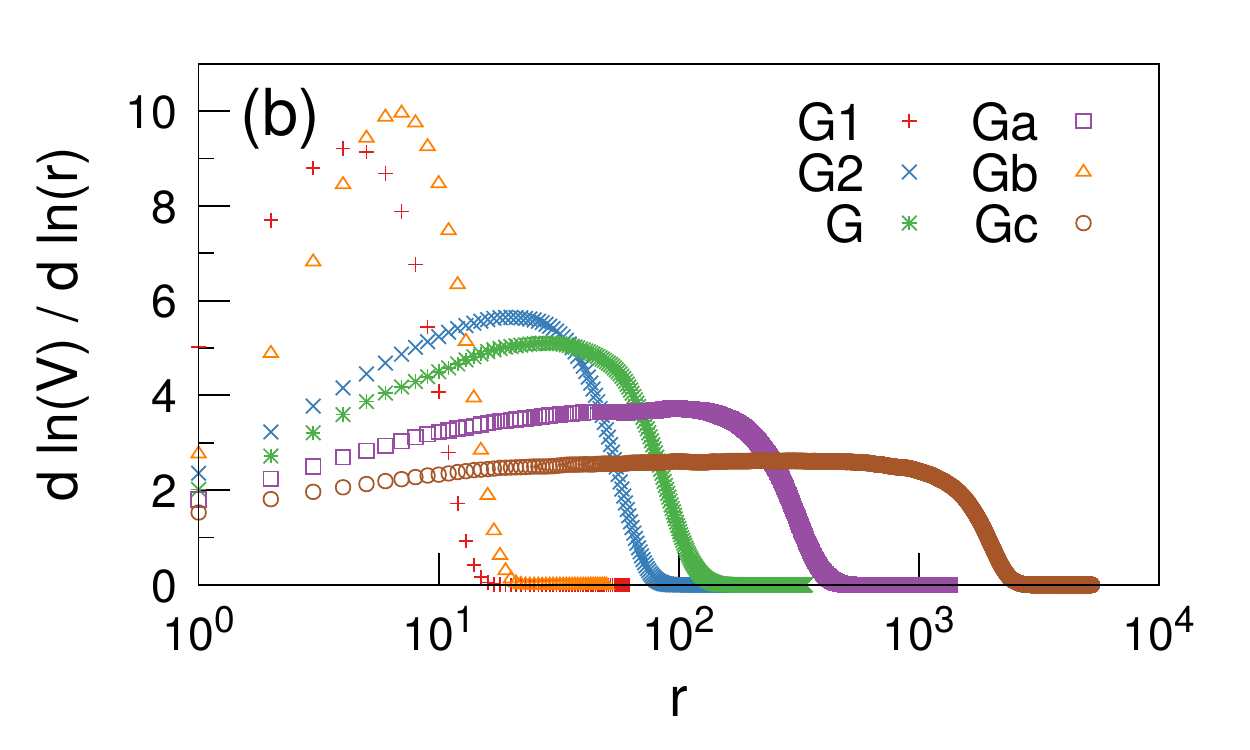}
\end{center}
\caption{
(a)  The average number of vertices $V(r)$ at a distance $r$ or smaller from a randomly chosen vertex as a function of $r$ for the models G, G2, Ga, Gb, and Gc of growing triangulation networks.  
(b) The logarithmic derivative $d\ln V(r)/d\ln r$. The presence of a plato on a curve for the logarithmic derivative indicates that the corresponding network is finite-dimensional. The networks in the numerical simulations are of $2^{28}$ vertices. The sizes of samples in the simulations for models G, Ga, Gb, and Gc are 16, 24, 16, and 16, respectively. For each sample, 100 vertices are chosen uniformly at random, around which the spheres of radius $r$ are made. 
}
\label{f12}       
\end{figure}

For a set of models of growing triangulations, we generated a number of realizations of $2^{28}$ vertices, and, using them, measured $V(r)$; see Fig.~\ref{f12}(a). At first sight, it seems from this figure that all these networks are finite dimensional, since the curves are visually close to a linear dependence in the log-log plot. However, inspecting the logarithmic derivative $d\ln V(r)/d\ln r$ shows that actually this is the case for only some of these networks, namely for the networks whose dependence $d\ln V(r)/d\ln r$ has a clear plateau. 
The difficulty is one needs a huge network to clearly observe a power-law dependence $V(r)$ in a wide range of $r$, say, several orders of magnitude. We observed the following: if we plot this curve for a larger network, this plateau is wider but its height is the same, $d\ln V(r)/d\ln r |_\text{plateau}=d_H$. On the other hand, if the dependence $d\ln V(r)/d\ln r$ has a peak instead of a plateau, then we face two possibilities. (i)   The peak in this dependence increases with network size $N$ up to infinity,  
%%(both the size and the width of the peak increase with $N$), 
which means $d_H=\infty$. 
(ii) The height of the peak stops growing when $N$ exceeds some value, 
%When $N$ is greater some value, the height of the peak stops growing 
while its width proceeds to increase, which means that this network is finite dimensional. Consequently, if we observe a peak in the dependence $d\ln V(r)/d\ln r$ for a given network, then $d_H$ is certainly greater than its maximum $d\ln V(r)/d\ln r |_\text{max}$. Based on these considerations, we concluded that our network has the space dimensions listed in Table~\ref{t2} [in the table we indicate $d_H=\infty$ when the peak $d\ln V(r)/d\ln r |_\text{max}$ exceeds, say, $10$ for our networks of $2^{28}$ vertices]. Note that while space dimension $4$ is well known as typical for random planar graphs \cite{ambjorn1997quantum}, any other finite value for a random triangulation based network is rather unexpected. 
Only model Ga has $d_H$ close to $4$. The table shows that while the models (G1, G2, GW) with small degree distribution exponent $\gamma$ have high or even infinite $d_H$, which is natural, the high or infinite $\gamma$ may be associated both with finite $d_H$ (models Ga and Gc) and with large or infinite $d_H$ (models G and Gb). 
The equilibrium networks E1, E2, and E3 that we generated in our numerical simulations were not sufficiently large to obtain $d_H$ confidently, so we have to leave three empty spaces in the table. 

\begin{figure}[t]
\begin{center}
\includegraphics[scale=0.52]{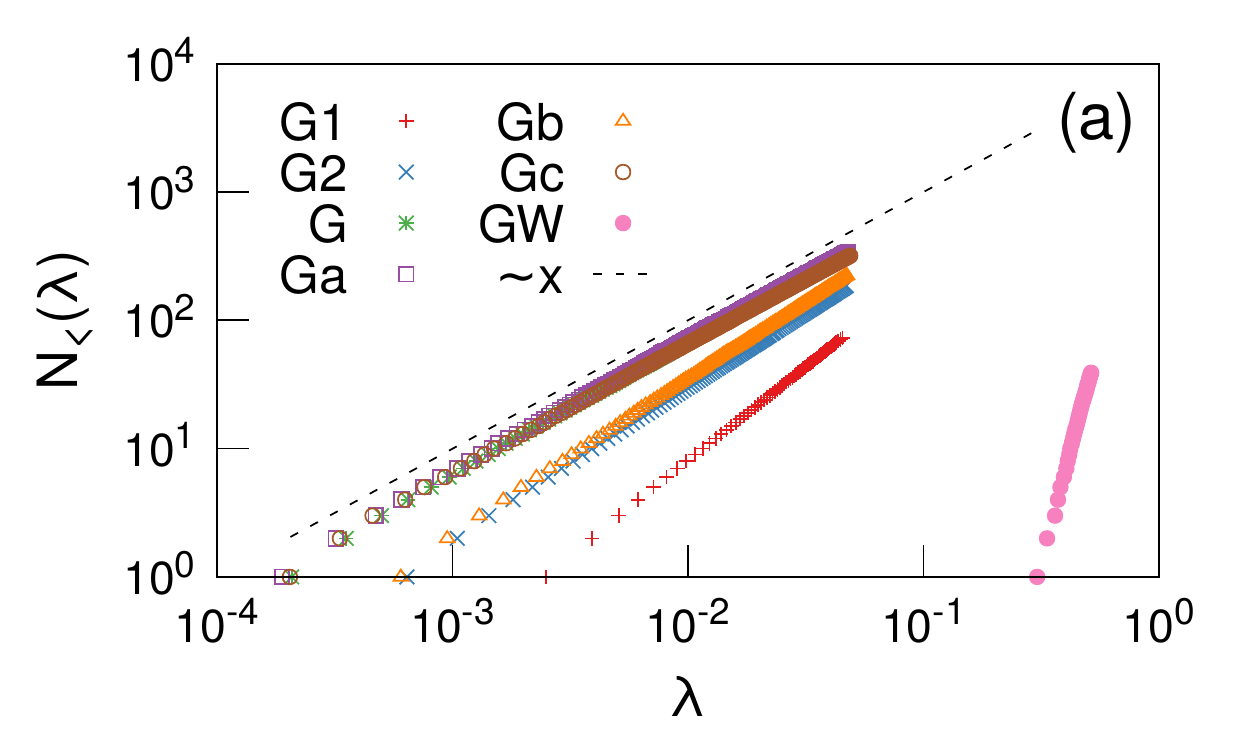}
\\[12pt]
\includegraphics[scale=0.52]{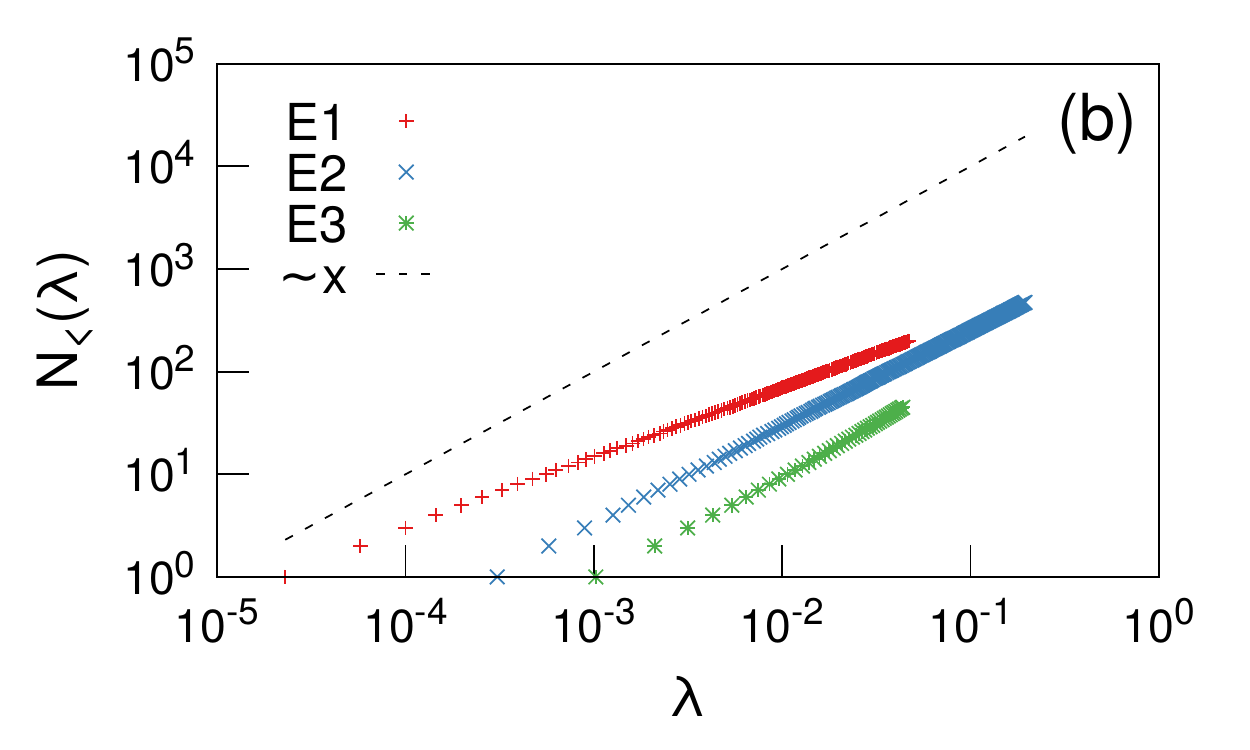}
\end{center}
\caption{ 
(a) Cumulative number of Laplacian spectrum eigenvalues, $N_<(\lambda) = \sum _{i:\,  0<\lambda_i\leq\lambda}1$, for the models G1, G2, G, Ga, Gb, Gc, and GW of growing triangulation networks. The spectra are obtained for the networks of $2^{17}$ vertices. The results are accumulated from 639, 256, 64, 32, 64, 32, and 32 samples for models G1, G2, G, Ga, Gb, Gc, and GW  respectively. 
%%The dashed line has the slope $1$, corresponding to the Laplacian dimension $d_L=2$. 
(b) 
%%Cumulative Laplacian eigenvalue density 
Cumulative number of Laplacian spectrum eigenvalues for the models E1, E2, and E3 of equilibrium triangulation networks. The spectra are obtained for the networks of $2^{15}$ vertices. For each model, the results are accumulated from 32 samples. 
%%(HMM... - IT SEEMS, FOR E1 --- LESS SAMPLES - ???). 
The dashed lines have the slope $1$, corresponding to the Laplacian dimension $d_L=2$. }
\label{f13}       
\end{figure}

Another key characteristic of the large scale organization of a network is its spectral dimension $d_S$, which, in simple terms, is its space dimension measured by using a 
%%random walk. 
diffusion process. For this process, the large time asymptotics of the density distribution at 
%%probability to find a random walker on 
%%return probability of a random walk to 
the starting vertex at time $t$ 
%%for an infinite network 
is 
\begin{equation}
p_0(t) \sim t^{-d_S/2}
%%.  
\label{160}
\end{equation}
for an infinite network. 
This corresponds to the following density of states of the Laplacian spectrum 
\begin{equation}
\rho(\lambda) \sim \lambda^{d_S/2-1}
%%.  
\label{170}
\end{equation}
for small eigenvalues $\lambda$. In finite networks, the dependence $p_0(t)$ has an exponential cutoff corresponding to the gap between the eigenvalue $0$ and the first nonzero eigenvalue of the Laplacian spectrum. For obtaining $d_S$ we inspected the number of (nonzero) eigenvalues in the Laplacian spectra smaller than $\lambda$, $N_<(\lambda) = \sum _{i:\,  0<\lambda_i\leq\lambda}1$, which is proportional to the cumulative density of states of the Laplacian spectra (has $N$ eigenvalues);   
%%the cumulative densities of states, $\rho_\text{cum}(\lambda)=(1/N)\sum_{i:\,  0<\lambda_i\leq\lambda}1/N$, in the Laplacian spectra of our models, 
see Fig.~\ref{f13}. 
To reduce fluctuations, which are large for small eigenvalues, we averaged eigenvalue $\lambda_i$ for each given $i$ over samples before calculating the cumulative numbers $N_<(\lambda)$. 
For each model we obtained a set of smallest eigenvalues of its Laplacian spectrum since only they were needed to find $d_S$. The curves in Fig.~\ref{f13} reach the value $N-1$ at large $\lambda$. 
%%NOTE THAT TO REDUCE FLUCTUATIONS, WHICH ARE 
%%PARTICULARLY 
%%LARGE FOR SMALL EIGENVALUES, 
%%VALUES OF $\lambda$, 
%%WE AVERAGED EIGENVALUE $\lambda_i$ FOR EACH GIVEN $i$ OVER SAMPLES BEFORE CALCULATING THE CUMULATIVE NUMBERS. 
The log-log plots in this figure provide the set of values of $d_S$ presented in Table~\ref{t2}. These resulting numbers differ strongly from the Hausdorff dimensions of these triangulation based networks and sit in the region from $1.4(2)$ for E1 to $2.9(2)$ for G1. 
%%(We will conclude in Sec.~\ref{s7} that network GW has $d_S=\infty$.) 
As one could expect, this region includes dimension $2$ of the simplex (triangle), although the observed deviations from this value are marked. Notice also that in these networks the spectral dimensions $d_S$ are finite even when $d_H$ is infinite, i.e., when they are small worlds.  
We shall suggest however based on our simulations in Sec.~\ref{s7} that the small-world network GW (model generating 
%%wormholes  
which play the role of long-range shortcuts) has its spectral dimension $d_S=\infty$ the same as $d_H$. 
The rest growing networks in Table~\ref{t2} have $d_H>d_S$.  

Notice the spectral dimension $d_S=1.4(3)$ for model E1. As we mentioned above, we did not obtain its Hausdorff dimension. We expect however that $d_H$ for this network is close to $2$. Interestingly, a similar combination of dimensions is valid for the ensemble of random connected trees (i.e., each member of the ensemble consists of a single connected component), in which $d_H=2$ and $d_S=4/3$ \cite{rammal1983random,jonsson1998spectral,durhuus2009hausdorff,wheater2009gravity}.

%%%%%%%%%%%%%%%%%%%%%%%%%%%%%%%%%%%%%%%%%%%%%%
%%%%%%%%%%%%%%%%%%%%%%%%%%%%%%%%%%%%%%%%%%%%%%

%%
%%

%%%%%%%%%%%%%%%%%%%%%%%%%%%%%%%%%%%%%%%%%%%%%%
%%%%%%%%%%%%%%%%%%%%%%%%%%%%%%%%%%%%%%%%%%%%%%

%%%%%%%%%%%%%%%%%%%%%%%%%%%%%%%%%%%%%%%%%%%%%%
%%%%%%%%%%%%%%%%%%%%%%%%%%%%%%%%%%%%%%%%%%%%%%

%%
%%

%%%%%%%%%%%%%%%%%%%%%%%%%%%%%%%%%%%%%%%%%%%%%%
%%%%%%%%%%%%%%%%%%%%%%%%%%%%%%%%%%%%%%%%%%%%%%

%%%%%%%%%%%%%%%%%%%%%%%%%%%%%%%%%%%%%%%%%%%%%%
%%%%%%%%%%%%%%%%%%%%%%%%%%%%%%%%%%%%%%%%%%%%%%

%%
\begin{figure}[t]
\begin{center}
\includegraphics[scale=0.52]{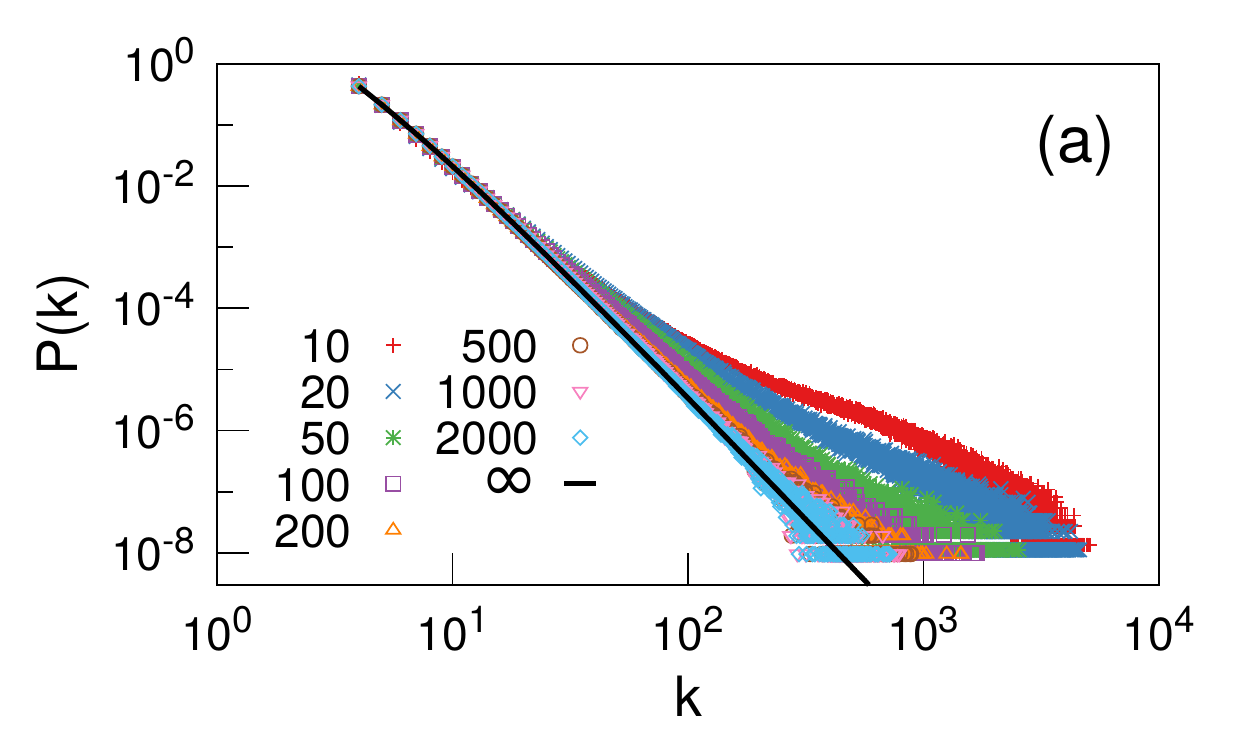} 
\\[12pt]
\includegraphics[scale=0.52]{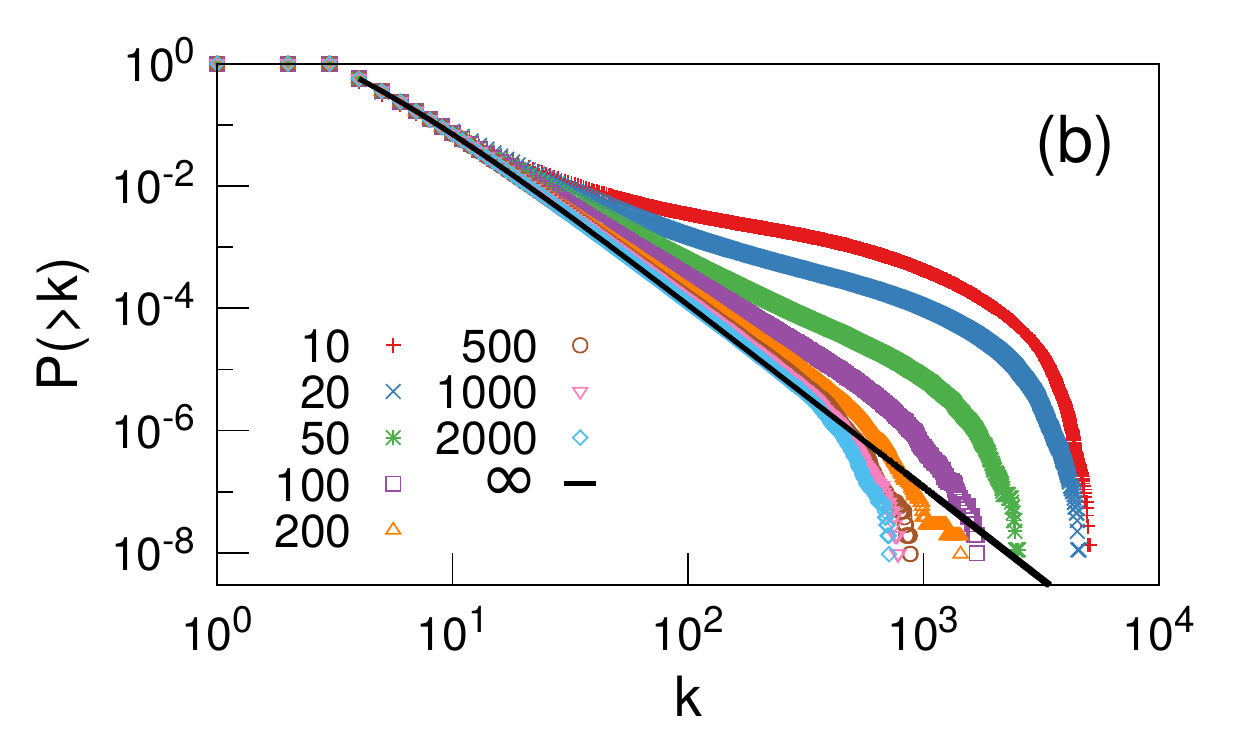}
\end{center}
\caption{
(a) Degree distribution of the model GW of a growing triangulation network with periodically merging triangles ($h$-holed torus). Different periods $\theta$ for introducing the 
%%wormholes is 
holes are considered, $\theta{=}10,20,50,100,200,500,1000,2000,$ and $\infty$. 
The resulting networks in the numerical simulations are of $2^{20}$ vertices, the averaging is over 10 (for $\theta=10$) or 100 (for $\theta\geq20$) samples. The infinite period $\theta$ provides model G2.   
(b) The corresponding cumulative degree distributions. 
}
\label{f14}       
\end{figure}
%%

%%%%%%%%%%%%%%%%%%%%%%%%%%%%%%%%%%%%%%%%%%%%%%
%%%%%%%%%%%%%%%%%%%%%%%%%%%%%%%%%%%%%%%%%%%%%%

%%%%%%%%%%%%%%%%%%%%%%%%%%%%%%%%%%%%%%%%%%%%%%
%%%%%%%%%%%%%%%%%%%%%%%%%%%%%%%%%%%%%%%%%%%%%%

%%
\begin{figure}[t]
\begin{center}
\includegraphics[scale=0.52]{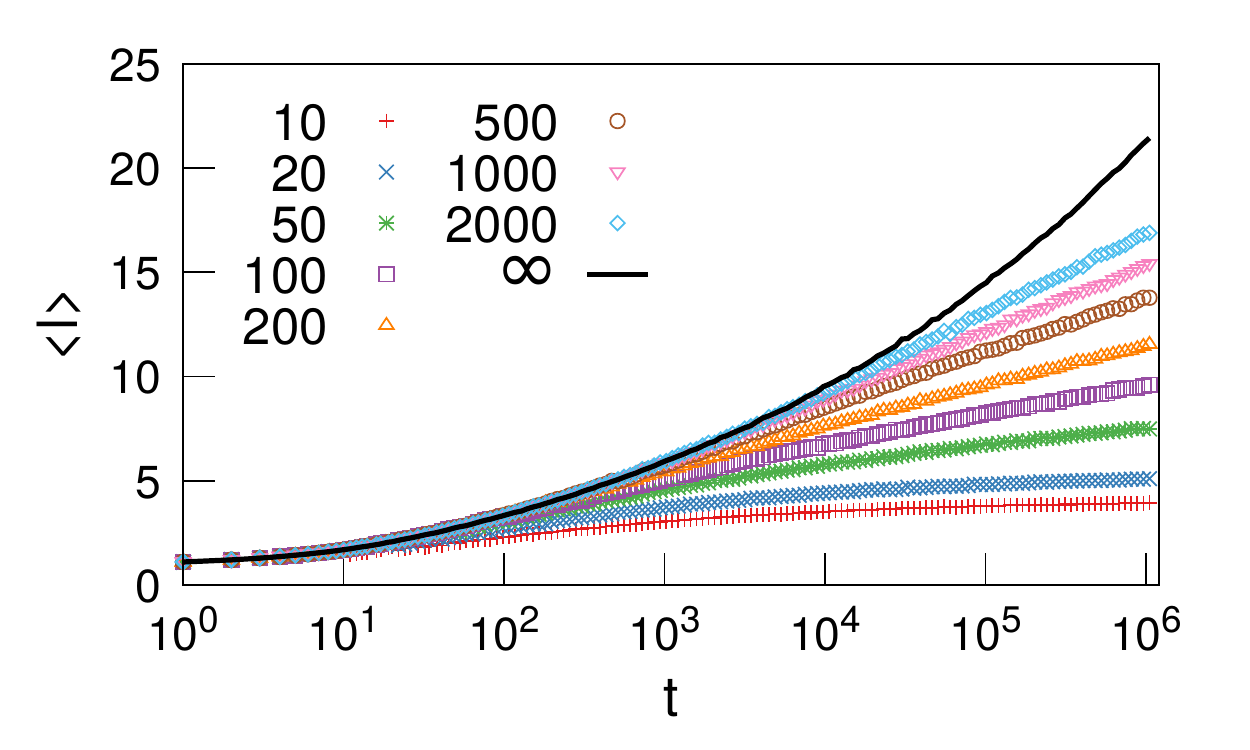}
\end{center}
\caption{
The time evolution of the average distance between vertices in the model GW of a growing triangulation network with periodically merging triangles. Different curves are obtained for different periods $\theta$ of introducing the holes, 
%%wormholes, 
$\theta{=}10,20,50,100,200,500,1000,2000,$ and $\infty$. The curve for $\theta\to\infty$ corresponds to model G2.
}
\label{f15}       
\end{figure}
%%

%%%%%%%%%%%%%%%%%%%%%%%%%%%%%%%%%%%%%%%%%%%%%%
%%%%%%%%%%%%%%%%%%%%%%%%%%%%%%%%%%%%%%%%%%%%%%

%%%%%%%%%%%%%%%%%%%%%%%%%%%%%%%%%%%%%%%%%%%%%%
%%%%%%%%%%%%%%%%%%%%%%%%%%%%%%%%%%%%%%%%%%%%%%
%%%%%%%%%.                         T A B L E  I I
%%%%%%%%%%%%%%%%%%%%%%%%%%%%%%%%%%%%%%%%%%%%%%
%%%%%%%%%%%%%%%%%%%%%%%%%%%%%%%%%%%%%%%%%%%%%%

\begin{table*}
\caption{
Key characteristics of network models of closed triangulations: exponent $\gamma$ of a degree distribution $P(k)\sim k^{-\gamma}$, Hausdorff dimension $d_H$, and  spectral dimension $d_S$. The values of $\gamma$ for models G1 and G2 are exact, the other numbers in the table were obtained from simulations. The simulated in this work networks E1, E2, and E3 were too small to obtain their Hausdorff dimensions. 
%%Here we mark exponent or dimension as infinite if our simulations .... 
%%INCLUDE OR NOT 
%%G$'$ 
%%(IT HAS $\gamma=\infty$, I AM NOT SURE ABOUT OTHER VALUES) - ? FOR Gb, SHOULD WE PUT $\gamma>4.5$ OR $\gamma=\infty$ - ? DIAMANTINO, HOW RELIABLE IS $\gamma=3.0(2)$ FOR E3? 
} 
%%$t_c$ is the critical point, $\beta$ and $\tau$ are the critical exponents of the percolation cluster and the finite cluster size zistribution, respectively, $f(0)$ is the critical amplitude, and shows}
%%\begin{center}
\begin{tabular}{ccccc}
\hline
\noalign{\smallskip}
%%& & & & 
%%\\
%%[-9pt]
model    
&   
\ \ exponent $\gamma$\  \ 
&  
\ \ Hausdorff dimension $d_H$\ \   
&  
\ \ spectral dimension $d_S$ 
 
\\[3pt]
\hline
%%\noalign{\smallskip}
%%\svhline
%%\noalign{\smallskip}
%%\hline
%%[0pt] 
%%& & & & 
\\[-11pt]
%%%%%\hline
%%%%%&&&&
%%%%%\\
%%\\
\multicolumn{4}{c}{Growing triangulations}
\\[3pt]
\hline
\\[-9pt]
G1   
& 
$3$ 
&$\ $  
%%0.92320750930(2) 
$\infty$
$\ $  & 
$2.9(2)$  
\\
G2  
& 
$4$ 
& 
${>}5.7$
& 
$2.4(2)$
\\
G  
& 
$\infty$  
&   
${>}5.0$
& 
$2.0(3)$  
\\
Ga 
& 
$\infty$ 
& 
$3.8(2)$
& 
$2.0(4)$
\\
Gb   
& 
${>}4.5$ 
&
$\infty$
& 
$2.4(3)$  
\\
Gc  
& 
$\infty$ 
& 
$2.6(2)$
& 
$2.0(3)$
\\[3pt]
\hline
\\[-11pt]
\multicolumn{4}{c}{Growing triangulation with increasing number of holes}
\\[3pt]
\hline
\\[-9pt]
GW   
& 
${\lesssim}2.0$ 
&  
$\infty$
& 
$\gtrsim 14$
%%$\infty$  
%%\\
\\[3pt]
\hline
%%\noalign{\smallskip}
%%\svhline
%%\noalign{\smallskip}
%%\hline
%%[0pt] 
%%& & & & 
\\[-11pt]
%%%%%\hline
%%%%%&&&&
%%%%%\\
%%\\
\multicolumn{4}{c}{Equilibrium triangulations}
\\[3pt]
\hline
\\[-9pt]
E1  
&
$\infty$  
& 
---
& 
$1.4(2)$
\\
E2   
&
$\infty$  
&$\ $  
%%0.92320750930(2) 
---
$\ $  & 
$1.9(4)$ 
\\
E3 
& 
$3.0(2)$ 
& 
---
& 
$2.1(2)$
\\
\noalign{\smallskip}\hline\noalign{\smallskip}
%%[2pt]
%%\hline
\end{tabular}
%%\end{center}
\label{t2}
\end{table*}

%%%%%%%%%%%%%%%%%%%%%%%%%%%%%%%%%%%%%%%%%%%%%%
%%%%%%%%%%%%%%%%%%%%%%%%%%%%%%%%%%%%%%%%%%%%%%
%%%%%%%%%%%%%%%%%%%%%%%%%%%%%%%%%%%%%%%%%%%%%%
%%%%%%%%%%%%%%%%%%%%%%%%%%%%%%%%%%%%%%%%%%%%%%

%%%%%%%%%%%%%%%
%%%%%%%%%%%%%%%%%%%%%%%%%%%%%%%%%%%
%%%%%%%%%%%%%%
%%%%%%%%%%%%%%%%%%%%%%%%%%%%%%%%%%%

\section{Generation of holes: 
%%wormholes: 
Evolving topology}
\label{s7}   

The closed manifolds considered above stayed homeomorphic to a sphere (or hypersphere) during the entire evolution. 
%%The models of the manifolds that we considered above had one common feature, namely during the entire evolution they stayed homeomorphic to sphere (or hypersphere). 
%%, that is their topology did not evolve. 
In contrast to these models, model GW from Table~\ref{t1} generates networks triangulating  manifolds with evolving topology. In this model two processes are applied in parallel. 
(i) At each step, the same move as in model G2 is made. 
(ii) In addition, at each $\theta$-th step, merging of two randomly chosen triangles (and annihilating these two faces as explained in Sec.~\ref{s4}) produces a 
%%wormhole 
hole (genus) in this manifold. 
The number of vertices in this network is $t(1-3/\theta)$ asymptotically. 
As a result we have an $h$-holed torus with a progressively growing number $h\cong t/\theta $ of holes.  
%%(wormholes in language of cosmology). 
%%Here we prefer to use the term ``wormholes'' to distinguish these topological features from ``holes'' in manifolds which one can cut from a manifold by a boundary.  
Model GW reduces to model G2 in the limit $\theta\to \infty$.  
%%Note that twisting of the merging triangles is not essential in our problems, since any ``twisted'' configuration can be untwisted by a series of Pachner moves or by applying operation S. 

Figure~\ref{f14} demonstrates the degree distributions and the cumulative degree distributions of the manifolds evolving according to rule GW.  The results were obtained by numerical simulations in which the networks were grown up to $2^{20}$ vertices for a set of periods $\theta$, from $10$ to $\infty$.  As is natural, for $\theta\to\infty$ we observe a power-law degree distribution of model G2, with exponent $\gamma=4$. For finite $\theta$, the degree distribution in the region of large degrees decays slower than in model G2. As $\theta$ decreases, this slow decay become observable at lower degrees, and it can be roughly estimated as $\sim k^{-2}$ or even slower. 

%%We suggest that this hump does not imply 
%%The sizes of the simulated networks were not sufficient to reliably conclude if this hump implies a condensation phenomenon. 

It is worthwhile to note that model GW is essentially similar to the aggregation growing network \cite{alava2005complex} (model D in the cited work) 
%%aggregation based network model D from Ref.~\cite{alava2005complex} 
in which, at each step, the end vertices of a uniformly randomly chosen edge merged together. The 
resulting 
aggregation process in Ref.~\cite{alava2005complex} was treated analytically. It was found that the network has a particularly slowly decaying degree distribution and, in a wide region of parameters, demonstrates a condensation phenomenon (a single vertex attracts a finite fraction of all connections). This is why the slowly decaying degree distributions of networks GW in Fig.~\ref{f14} are not surprising. However, in contrast to Ref.~\cite{alava2005complex}, we did not observe condensation phenomena in model GW. One of possible reasons for that is the additional constraint that forbids merging first- and second- nearest-neighboring triangles. This also makes an analytical treatment of the model more challenging than in Ref.~\cite{alava2005complex}. 
%THIS MODEL IS SIMILAR TO THE GROWING NETWORK WITH MERGING OF VERTICES WHICH WE CONSIDERED WITH MIKKO ALAVA IN 2005 (PRE) \cite{alava2005complex}. IN THAT MODEL WE OBSERVED CONDENSATION. IN FACT, MODEL GW CAN BE RATHER EASILY SOLVED (APPROXIMATELY ! --- NEGLECTING CORRELATIONS !) ANALYTICALLY, SEE THAT OUR OLD PAPER.  
%THE COMPLICATION IS THAT IN GW WE HAVE AN ADDITIONS CONSTRAINT. 1ST- AND 2ND-NEAREST NEIGHBORING TRIANGLES CANNOT MERGE. THIS CONSTRAINT HAMPERS THE EMERGENCE OF CONDENSATION PHENOMENON, WHICH WAS POSSIBLE IN THE MODEL OF REF. \cite{alava2005complex}.

In model GW, hubs preferentially (proportionally to degree, i.e., local curvature) participate in emergence of %%wormholes. 
holes. On the other hand, the birth of a hole 
%%wormhole 
produces vertices of higher degrees. So hubs and 
holes 
%%wormholes 
co-evolve and strongly correlate with each other. 
%%Triangles attached to vertices of high degree (hubs) have 
Furthermore, for sufficiently small values of the merging period $\theta$, in particular for $\theta=10$, we observed that in a fraction of runs, the evolution process GW stalled during the observation time, which was  about $10^6$ steps, and we had to restart the process from zero. The reason for this stall is that this network is so compact (the average separation of vertices approaches only $4$ at $t\sim 10^6$; see Fig.~\ref{f15}) that it is possible that at some instant our algorithm cannot find triangles relevant for merging.  (Recall that they cannot be first- and second-nearest neighbors by the rules of the model).

%%VERTICES OF HIGH DEGREES ARE ASSOCIATED WITH WORMHOLES AND CO-EVOLVE. 

%%NOTE THAT FOR $\theta=10$, THE EVOLUTION OF SOME SAMPLES WAS STACKED AT SOME MOMENT BEFORE $t=10^6$. IF THIS HAPPEN, WE RESTART THE PROCESS FROM ZERO. 

To characterize long-range properties of these manifolds we inspected the evolution of the average distance $\langle\ell\rangle$ between vertices of the generated networks for different values of parameter $\theta$; see Fig.~\ref{f15}. 
%%, $\langle\ell\rangle$ vs $t$, number of steps of the process, playing the role of a current time. 
The curve for $\theta=\infty$ increases with $t$ in this range, $1\leq t\leq 10^6$,  faster than the first power of the logarithm. 
Our more thorough analysis in Sec.~\ref{s6} showed that network G2 has very large, probably infinite, dimension $d_H$. 
%%$\ln t$. 
%%COMPARE THE CURVE FOR $\theta=\infty$ WITH THE ANALYSIS OF THE CURVE FOR G2 OBTAINED BY ANOTHER METHOD 
In this sense, network G2 already can be called a ``small world''. Nonetheless, as Fig.~\ref{f15} demonstrates, the additional merging of triangles during the evolution produces a particularly strong small-world effect, i.e., the network becomes even more compact. For $\theta<1000$, $\langle\ell\rangle$ increases with $t$ even slower than the first power of logarithm.  The emerging 
holes 
%%wormholes 
play the role of shortcuts between random vertices of the Watts-Strogatz model of a small-world network \cite{watts1998collective} but applied not to a one-dimensional lattice, as in the original model, but to a very high-dimensional  or even infinite-dimensional network. So, according to Fig.~\ref{f15}, we may arrive even 
%%The slowly increasing curves $\langle\ell\rangle(t)$ for $\theta<1000$ in Fig.~\ref{f15} supposes that this results in what was 
at so-called ``ultrasmall worlds'' \cite{cohen2002ultra,dorogovtsev2003metric}. 
%%Notice that at $\theta=10$, the average separation of vertices in this network is only $4$.  
%%G2 IS INFINITE DIMENSIONAL BUT ITS DIAMETER FOR THIS RANGE OF SIZES GROWS FASTER THAN LOG, SOMETHING OF KIND POWER OF LOG (IT SEEMS, IT  WILL ACTUALLY APPROACH AN ORDINARY LOG AT VERY LARGE SIZES WHICH WE CANNOT OBSERVE). 

%%THE WORMHOLES GENERATE AN ADDITIONAL SMALL-WORLD EFFECT, SO THE FIGURE SHOWS SIGNIFICANT REDUCTION OF THE DIAMETER. IN THIS RANGE OF SAMPLE SIZES, THE RESULTING DIAMETER INCREASES WITH SIZE EVEN SLOWER THAN LOG IF $\theta<100$. (MAYBE, IT  WILL ACTUALLY APPROACH AN ORDINARY LOG AT VERY LARGE SIZES WHICH WE CANNOT OBSERVE - ??). 

\begin{figure}[t]
\begin{center}
\includegraphics[scale=0.52]{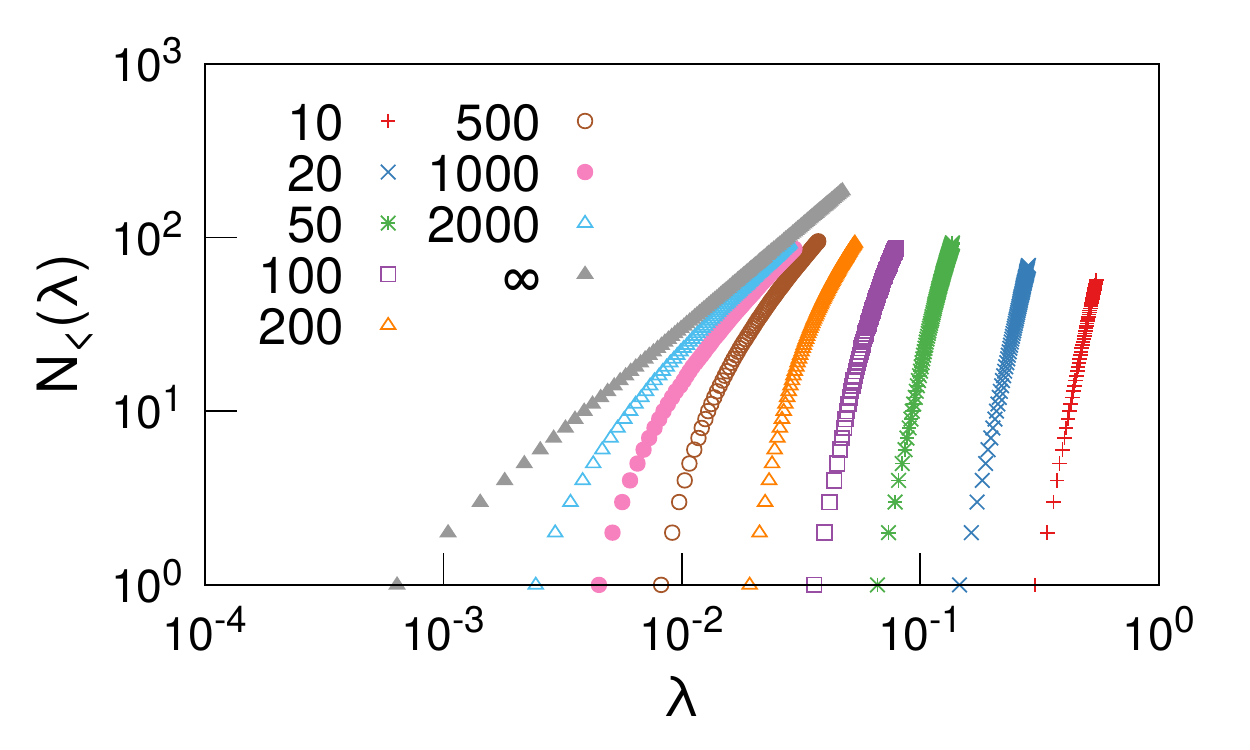}
\end{center}
\caption{
Cumulative number of Laplacian spectrum eigenvalues of model GW for a set of values $\theta{=}10,20,50,100,200,500,1000,2000,$ and $\infty$. The network size is $2^{17}$ vertices. 
}
\label{f16}       
\end{figure}

\begin{figure}[t]
\begin{center}
\includegraphics[scale=0.52]{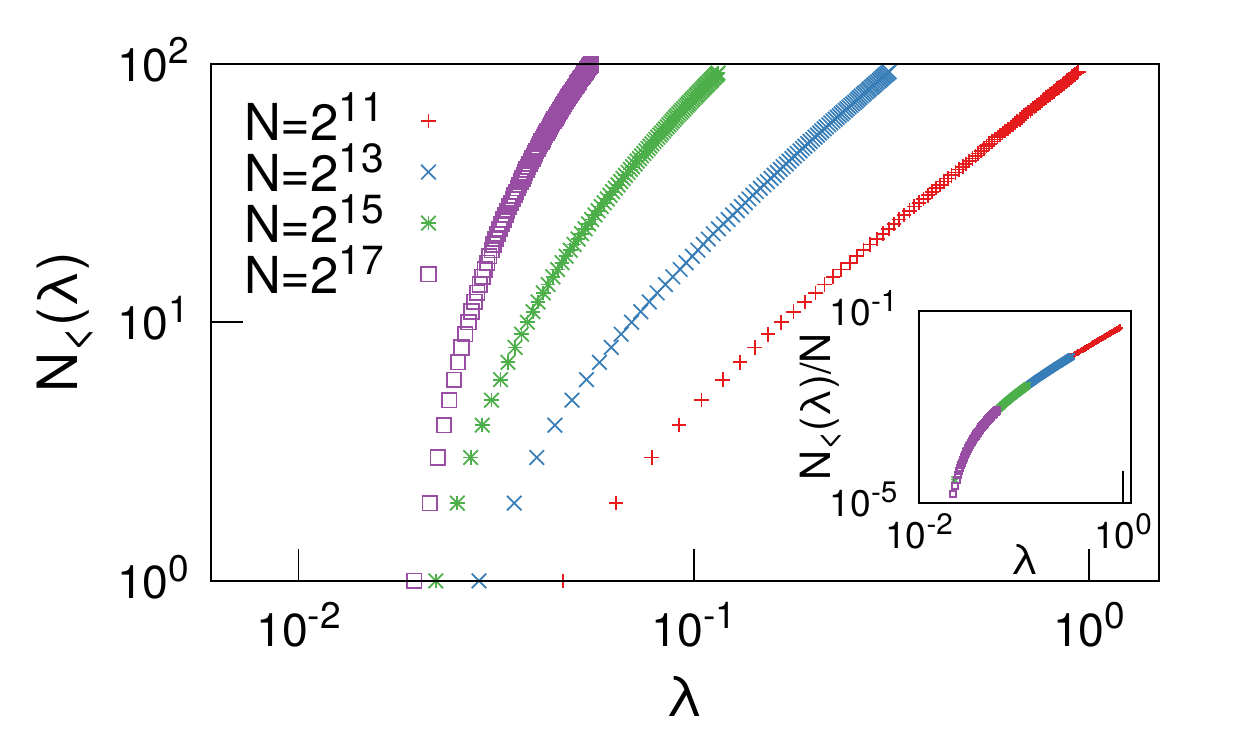}
\end{center}
\caption{
Cumulative number of Laplacian spectrum eigenvalues of model GW, $\theta{=}200$, for a set of network sizes. 
%%Notice the size effect for the slope of the curve at small $\lambda$. 
The inset shows the cumulative density of the Laplacian spectrum $N_<(\lambda)/N$ vs $\lambda$ for the same set of network sizes. Notice that all the curves in the inset precisely coincide in the respective regions of $\lambda$.
}
\label{f17}       
\end{figure}

Finally we investigated the Laplacian spectrum of model GW to obtain its spectral dimension $d_S$. Figure~\ref{f16}, showing the cumulative Laplacian spectra of the networks with different $\theta$, demonstrate that $d_S$ is very high if not infinite. For example, for the network with $\theta=10$, see Figs.~\ref{f13} and \ref{f16}, we observe the slope about $14$ of the curve in the log-log plot for small $\lambda$. Interestingly, in Fig.~\ref{f16}, this slope saturates as period $\theta$ becomes smaller than $200$.  However, the region of $\lambda$, where this power-law asymptotic is observed, is very narrow. 
We suggest that this slope (and the value of $d_S$) should be even infinite for the infinite networks generated by this model. 
The difficulty is that the investigated networks are rather small, $10^4$---$10^5$ vertices, and, what is even more important, are very compact, the mean separation of vertices is about $4$ when $\theta=10$, while the size effect is strong, see Fig.~\ref{f17} showing the dependencies $N_<(\lambda)$ at $\theta=200$ for different network sizes $N$. This figure demonstrates that the slope of the cumulative Laplacian spectrum in the region of small $\lambda$ increases with network size. For a fixed $\theta$, the number of 
%%wormholes-- 
holes (shortcuts) is proportional to the network size. So this plot also describes the effect of shortcuts on the Laplacian spectrum of this specific growing small-world network. Notice that for the cumulative density of states in the Laplacian spectrum, $N_<(\lambda)/N$, the curves obtained from Fig.~\ref{f17} practically coincide with each other in the respective regions of $\lambda$; see the inset of Fig.~\ref{f17}. 
%%So the size effect on the Laplacian spectrum is 

%%%%%%%%%%%%%%%%%%%%%%%%%%%%%%%%%%%%%%%%%%%%%%
%%%%%%%%%%%%%%%%%%%%%%%%%%%%%%%%%%%%%%%%%%%%%%

%%
%%

%%%%%%%%%%%%%%%%%%%%%%%%%%%%%%%%%%%%%%%%%%%%%%
%%%%%%%%%%%%%%%%%%%%%%%%%%%%%%%%%%%%%%%%%%%%%%

%%%%%%%%%%%%%%%%%%%%%%%%%%%%%%%%%%%%%%%%%%%%%%
%%%%%%%%%%%%%%%%%%%%%%%%%%%%%%%%%%%%%%%%%%%%%%

%%
%%

%%%%%%%%%%%%%%%%%%%%%%%%%%%%%%%%%%%%%%%%%%%%%%
%%%%%%%%%%%%%%%%%%%%%%%%%%%%%%%%%%%%%%%%%%%%%%

%%%%%%%%%%%%%%%
%%%%%%%%%%
%%%%%%%%%%%%%%
%%%%%%%%%%%

\section{Discussions and conclusions}
\label{s8}

%%In our work we 
We did not 
consider 
%%touch upon 
%%a number of 
some issues 
%%, essential part of which are 
related to an interplay between topology, metric structure, and the nonequilibrium nature of many of our evolution models. One of these interesting issues is the possible twisting of merging triangles. This twisting would mean that the shortest path between two points may spiral around a 
%%wormhole, 
hole, 
i.e., it relates to a metric structure.  We did not consider this twisting in Sec.~\ref{s4} since a (long) sequence of Pachner moves (or operation $S$ from Sec.~\ref{s3}) can smoothly untwist a configuration of this kind. This possibility of untwisting should be typical for equilibrium models, but the situation may be more complicated for nonequilibrium networks of this kind. Indeed, if we introduce twisting for each of the frequently occurring mergings into a growing network of the GW kind, then the second channel of the evolution process will have no time to complete untwisting. Analysis of the interesting resulting object is beyond the scope of the present work. 
 
%%AT FIRST SIGHT, THE twisting of the merging triangles is not essential in our problems, since any ``twisted'' configuration can be untwisted by a series of Pachner moves or by applying operation S. SO, IT IS VERY UNDERSTANDABLE THAT THERE CANNOT BE INTRODUCED ANY TWISTING NUMBER FOR TRIANGULATIONS OF HOLED CLOSED SURFACES --- SMOOTH UNTWISTING IS ALWAYS POSSIBLE. HOWEVER, IT SEEMS, THIS ARGUMENT ACTUALLY WORKS FOR EQUILIBRIUM MODELS OF THIS KIND

%%IF THE GROWING MODEL, YOU CAN INTRODUCE NEW TWISTING BY MERGING TWISTED TRIANGLES MUST FASTER THAN ANY UNTWISTING BY LONG CHAINS OF PACHNER OR S. SO IN NON-EQUILIBRIUM SITUATIONS, TWISTING IS, IN PRINCIPLE, POSSIBLE. WE DO NOT CONSIDER THIS SITUATION. 

We explained that for precise measuring space dimensions of our objects one should simulate very large networks. More extensive simulations than in this work can provide $d_H$ values even for our equilibrium models, for which we succeed to reach only sizes of $10^5$ vertices, not sufficient to obtain $d_H$.  

%%FUTURE WORK: 

%%OUR SIMULATIONS FOR EQUILIBRIUM MANIFOLDS ARE FOR TOO SMALL NETWORKS TO FIND THEIR SPACE DIMENSIONS --- $d_H$ OF EQUILIBRIUM MANIFOLDS 

We derived analytical expressions for degree distributions of a few our models. We suggest that an analytical, at least approximate, theory of a much wider range of these models is, in principle, doable, in particular for model GW, following ideas from Ref.~\cite{alava2005complex}. 
Furthermore, the rules of our evolution models do not specify edge lengths. We leave 
more detailed 
models for simplicial complexes with edges of different length for future work.  

It is worthwhile to mention that in the complex networks literature the term ``network topology'' is interpreted  frequently as a structural organization, a global structure, and so on. In contrast to this, in the present article we treat this term in the standard mathematical sense and consider the full set of topological features of our systems.

%%SOME OF EQUILIBRIUM MODELS ALLOW ANALYTICAL RESULTS FOR DEGREE DISTRIBUTIONS - WE OBTAINED ONLY FOR GROWING 

%%ANALYTICAL THEORY FOR MODELS A LA GW. 

The zoo of networks that we considered is just the tip of the iceberg. We focused on evolution processes not related to boundaries.  
%%but, in particular, 
However merging of simplices and generation of 
%%wormholes 
holes can be considered not only in closed manifolds, as in Sec.~\ref{s7}, but also in manifolds with boundaries. Moreover, the number of boundaries 
%%(``holes'') 
can also be evolving; 
new boundaries 
%%these 
%%holes 
can be generated progressively. 
%%, and so on. 
Other promising generalizations and variations of our models are also possible. In particular, if instead of the annihilation of two merging faces in the rules of model GW, we introduce merging two triangular faces into one, we shall obtain an expanding foam. 

We found that the constraint that a network is a triangulation produces a strong difference from typical planar graphs, 
random geometric graphs, and other networks embedded in metric spaces \cite{ambjorn1997quantum,penrose2003random,dall2002random,daqing2011dimension}. This difference lays beyond such local characteristics as degree distributions. Our models provide a rich array of dimensions of generated metric spaces and evolving topologies with a varying set of topological features. We obtained a wide spectrum of Hausdorff and spectral dimensions for our models of evolving triangulations. Notably, for some of them, $d_H$ is infinite, while $d_S$ is finite, see Table~\ref{t2}. %%In this case, it takes a very long time for a random walker to investigate a network despite its extremal compactness 
In this situation, 
%%a random walk looks like 
diffusion processes on a network look like in a finite-dimensional metric space despite the extremal compactness of the network. 

We observed that ``physical'' 
%%complex 
stochastic network models used for interpretation of evolving simplicial complexes produce a set of surprising results for their local properties (heavy tailed degree distributions) and global ones (unusual values of space dimensions, topological features, 
%%wormholes, 
holes, coupled with high local curvature). These effects combine the evolution, topology, and geometry of the considered objects. Although our conclusions  are made for abstract mathematical structures, we suggest that more detailed versions of our models and algorithms could be applied to real-world systems and processes. 
Triangulations are in the very heart of modern civilization providing the main method of treatment of surfaces in topography, engineering, hydrodynamics and aerodynamics, visualization techniques, and everywhere. 
We 
%%also 
suggest that an important application of our work could be development of efficient stochastic algorithms generating triangulations and higher-dimensional simplicial complexes with desired characteristics and features.

%%DEAR ALL: ANY IDEA FOR A GREAT LAST SENTENCE - !!!???

%%%%%%%%%%%%%%%%%%%%%%%%%%%%%%%%%%%%%%%%%%%%%%%%%%%%%%%%%%%%%%%%%%
%%%%%%%%%%%%%%%%%%%%%%%%%%%%%%%%%%%%%%%%%%%%%%%%%%%%%%%%%%%%%%%%%%

\begin{acknowledgments}
%%This work was partially supported by ??? 
%%the FET IP Project MULTIPLEX 317532.
%%, and also by the SOCIALNETS EU project. 
D.~C.~S. was supported by FEDER Grant No. POCI-01-0145-FEDER-007688. R.~A.~C. acknowledges  Grants No. BPD-35/I3N/SET2016/21537 and No. FCT SFRH/BPD/123077/2016. 
\end{acknowledgments}
%%%
%%%

%%%%\bibliography{evolving_manifolds}

%merlin.mbs apsrev4-1.bst 2010-07-25 4.21a (PWD, AO, DPC) hacked
%Control: key (0)
%Control: author (0) dotless jnrlst
%Control: editor formatted (1) identically to author
%Control: production of article title (0) allowed
%Control: page (1) range
%Control: year (0) verbatim
%Control: production of eprint (0) enabled
%

\end{document}